# Observation of the Josephson effect in superhydrides: DC SQUID based on (La,Ce)$H_{10+x}$ with operating temperature of 179 K

Dmitrii V. Semenok[1,†,*], Ivan A. Troyan[2,†], Di Zhou[1,†,*], Wuhao Chen[3,4] , Ho-kwang Mao [1] and Viktor V. Struzhkin[1,5,*]

[1] Center for High Pressure Science & Technology Advanced Research; Bldg. #8E, ZPark, 10 Xibeiwang East Rd, Haidian District, Beijing, 100193, China

[2]«A.V. Shubnikov Institute of Crystallography» of the Kurchatov Complex of Crystallography and Photonics (KKKiF); 59 LeninskyProspekt, Moscow, 119333, Russia

[3]Quantum Science Center of Guangdong–Hong Kong–Macao Greater Bay Area (Guangdong); No. 3, Binlang Road, Futian District, Shenzhen, China

[4]Department of Physics, Southern University of Science and Technology; No. 1088, Xueyuan Avenue, Nanshan District, Shenzhen, 518055, China

[5]Shanghai Key Laboratory of Material Frontiers Research in Extreme Environments (MFree), Shanghai Advanced Research in Physical Sciences (SHARPS); Pudong, Shanghai 201203, China

*Corresponding authors: Dmitrii Semenok (dmitrii.semenok@hpstar.ac.cn), Di Zhou (di.zhou@hpstar.ac.cn), and Viktor Struzhkin (viktor.struzhkin@hpstar.ac.cn)

†These authors contributed equally to this work

**Abstract:** Among known materials, hydride superconductors have the highest critical temperatures and are very promising as a basis for electronic sensors. Superconducting quantum interference device (SQUID), due to its unique sensitivity to magnetic fields, is the most important application of superconductors in microelectronics. In this work, we describe a direct current SQUID made of lanthanum-cerium superhydride (La, Ce)$H_{10+x}$ ($x$ is between –1 and +2) at pressure of 148 GPa, with operating temperature of 179 K and bias current of about 2 mA. When placing (La, Ce)$H_{10+x}$ in a modulated magnetic field (0.1-0.005 Hz, 5 G), we observed generation of higher harmonics up to 18$\nu_0$ and a periodic dependence of the sample resistance on the magnetic flux density $R \propto \sin(\pi\Phi/\Phi_0)$. We demonstrate that the (La, Ce)$H_{10+x}$ SQUID with a size of ~ 6 μm, operates in the mode of low thermal fluctuations and can be used to detect magnetic fields below 0.1 G. Our findings pave the road to more advanced applications of the Josephson effect and SQUIDs made of hydride superconductors.

**Graphical abstract**

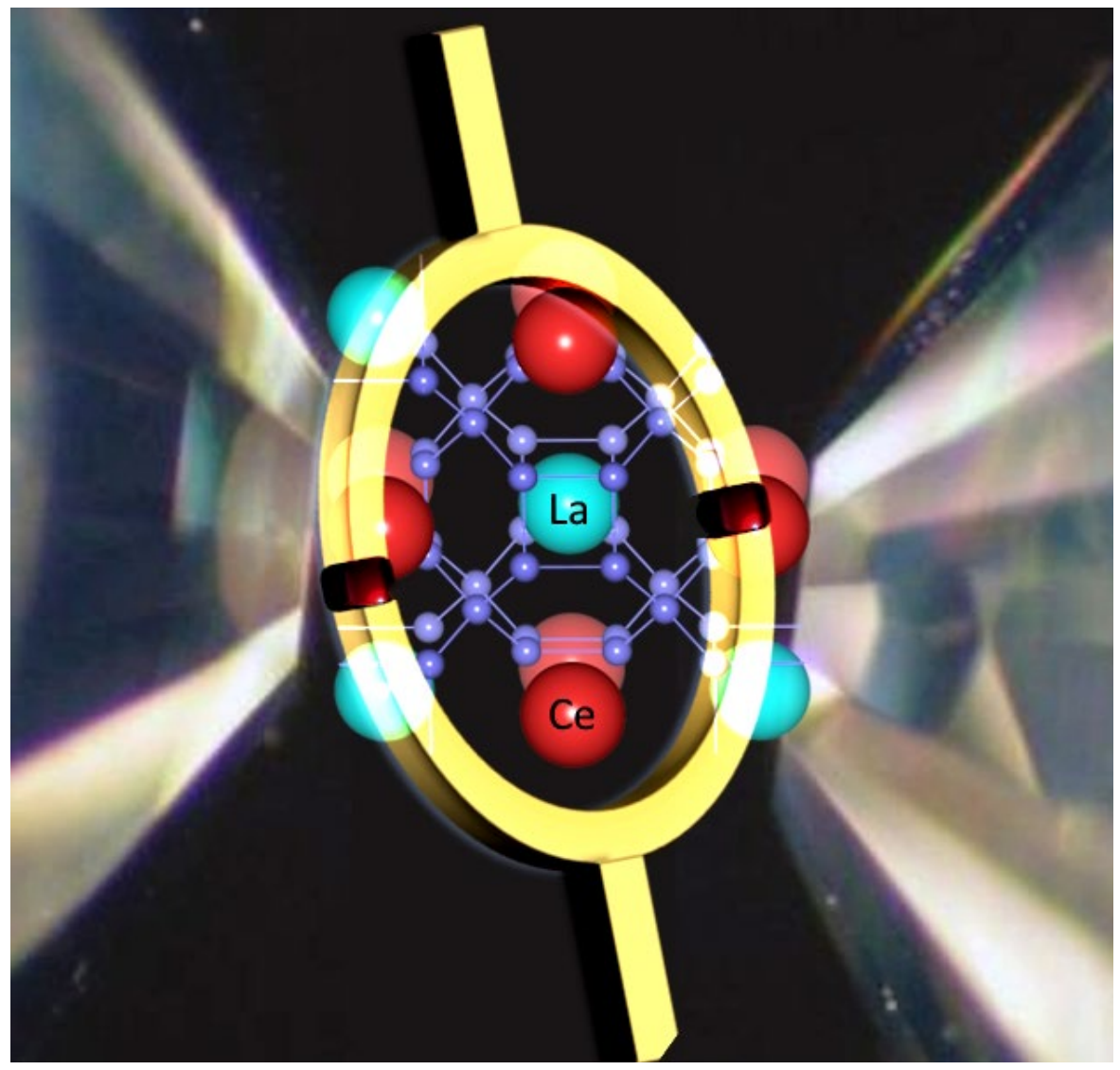

## Introduction

SQUID is a superconducting (SC) quantum interference device based on superconducting loops containing at least one Josephson junction [1,2]. Practically, it can be made from rings of a SC material with one or several weak links where the critical transition temperature ($T_C$) is lower than in the rest part of ring. Such device turns to be a very sensitive magnetometer that can be used to measure extremely weak magnetic fields as low as $5\times10^{-14}$ T [3,4]. SQUIDs can be designed on the basis of pure Nb or Al with oxide insulating Josephson junctions or weak links made of single walled carbon nanotubes [5,6]. To maintain superconductivity, the entire device needs to operate under very low temperature, cooled with liquid helium. High-temperature SQUID sensors based on cuprates (e.g., YBCO) are cooled by liquid nitrogen which is, from a practical point of view, easier than liquid helium cooling, but higher operating temperature leads to higher level of noise and less sensitivity than conventional He-temperature SQUIDs [7,8]. Despite this, high-temperature SQUID sensors have many applications in biology including magnetoencephalography [9], magnetic marker monitoring [10], cardiology, magnetic resonance imaging in low fields [11]. Reduced insulation thicknesses necessary for liquid nitrogen cooling enable the devices to be placed much closer to the sample under investigation, thus boosting the magnetic signal level. As we will see below, in some high-pressure diamond anvil cells, the sample and the SQUID sensor are separated by a physical barrier of only 5-6 mm, which is overcome by introducing inductive coupling using additional normal metal circuits.

High-temperature SQUID sensors make it possible to detect weak magnetic fields of various organs (brain, stomach, heart), to study their electrical activity and its spatial distribution. Many SQUID applications are associated with physical research, for example, the creation of quantum superconducting qubits and computers based on them [12], a scanning SQUID microscope [13], the study of the magnetic properties of materials, the search for the quadrupole moment of the electron [14], to make cryogenic particle detectors [15] and in many other areas. As the history of SQUID technology shows, both types of SQUIDs – noisy high-$T_C$ devices (based, for example, on cuprates) and precise low-noise low-$T_C$ magnetometers have occupied their independent niches in the application market.

Problem of increasing the operating temperature of high-$T_C$ SQUIDs rests on the search for new superconducting materials with a higher $T_C$. Fortunately, a kind of "hydride revolution" in superconductivity began in 2014, leading to the discovery of high-$T_C$ superhydrides such as $H_3S$ [16], $LaH_{10}$ [17,18], which has a current record of $T_C$ = 250 K, $YH_6$ [19,20], $ThH_{10}$ [21], $CeH_9$ [22], $SnH_4$ [23], (La,Ce)$H_{9-10}$ [24,25] and many others. The last one, lanthanum-cerium polyhydride, is one of the most convenient samples for high-pressure experiments: due to the stabilizing effect of Ce, its critical temperature $T_C$ = 190-200 K is achieved already at a pressure of 100-120 GPa [24,25]. Such properties allow to work with large-sized samples and diamond anvils with a diameter of 100-150 µm. It was long thought that due to the microscopic size, the need to use diamond anvils and high-pressure anvil cells (DACs), high-temperature superconductivity in hydrides would not find practical applications.

In this work, we show that this is not the case, and already existing sample of La-Ce superhydride in a DAC demonstrates a pronounced direct current (DC) SQUID effect and can be used as a sensor of weak magnetic fields. It is conceivable that optimization of sample geometry by using microfabrication tools could lead to substantial improvement in SQUID sensitivity and control, thus opening many application areas for superhydrides.

## Results

### *DC measurements*

High-pressure electrical diamond anvil cell DAC H with a hydride sample synthesized from a $La_{80}Ce_{20}$ alloy and ammonia borane was used in our experiments (Fig. 1A). A sharp drop in electrical resistance at $T_C$(onset) = 200-205 K was observed in both heating and cooling cycles using DC delta mode (Fig. 1B) and alternating current (AC) measuring circuits (Fig. 3A). Considering that the found $T_C$ is higher than it was determined in early works for cubic and hexagonal (La, Ce)$H_{9+x}$ [24,25], and also results of X-ray diffraction analysis of a hexagonal (La,Ce)$H_{12}$ sample prepared in the same manner from $La_{86}Ce_{14}$ alloy and $NH_3BH_3$ (Figs. 4D-F), we will further attribute the chemical formula (La, Ce)$H_{10+x}$ to the studied compound (similar to $LaH_{10+x}$ [26]).

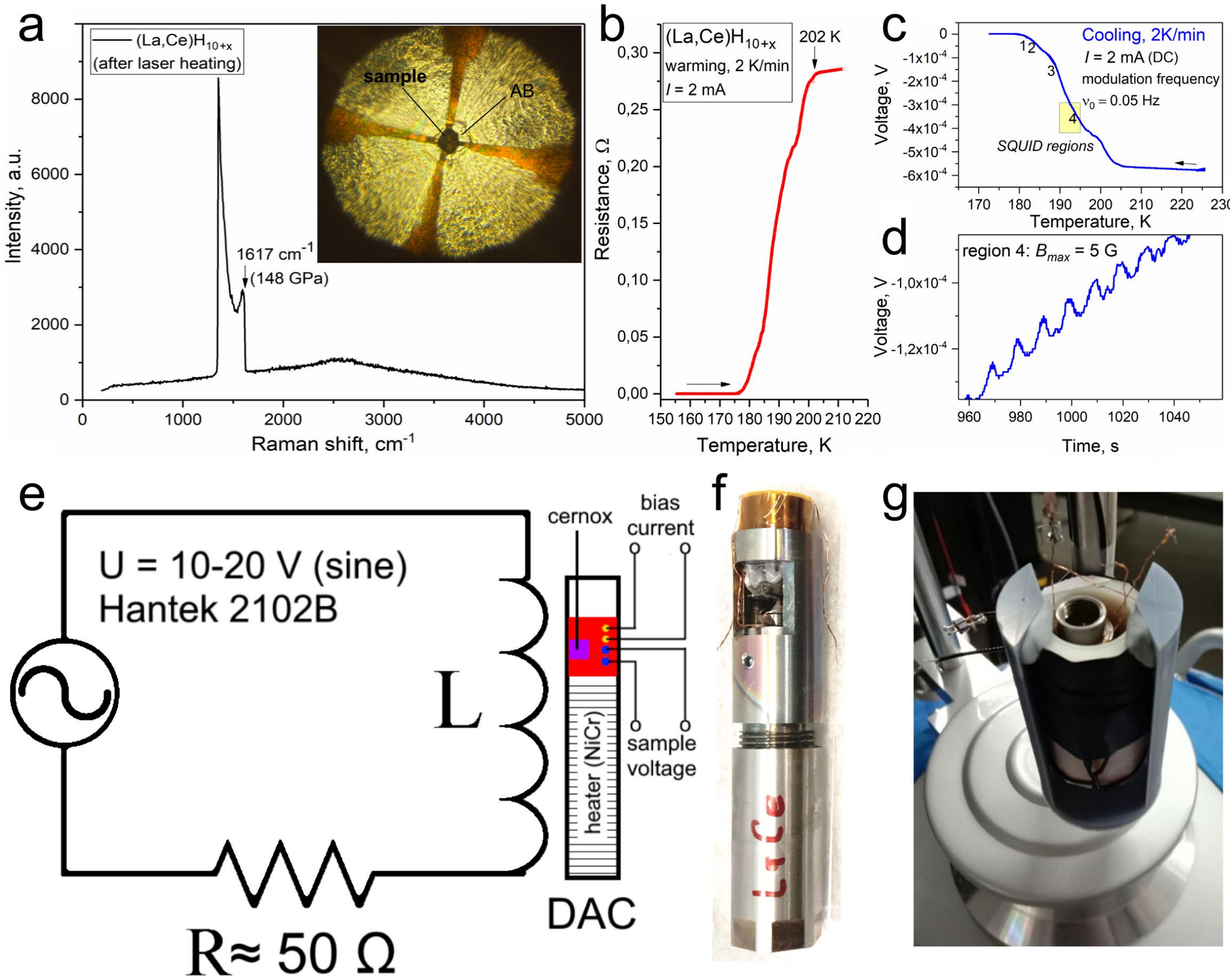

**Fig. 1. Properties of the (La, Ce)$H_{10+x}$ sample in the DAC H and the scheme of measurements.** (**A**) Raman spectrum of the sample after laser heating. The edge of the diamond Raman signal corresponds to a pressure of 148 GPa [27]. Inset: photo of the DAC's culet with compressed sample. (**B**) Superconducting transition in (La, Ce)$H_{10+x}$ observed within the warming cycle. (**C**) Cooling of a (La, Ce)$H_{10+x}$ sample in a magnetic field of ~ 5 G modulated by a low frequency of 0.05 Hz. To measure the voltage drop across the sample, we used a DC current of 2 mA. The modulation in the sample resistance appears only in a few temperature intervals: around 187 K, 182-184 K, 181-179 K etc. The most convenient interval for studying the SQUID effect is in the region of the lowest resistance around 179 K. (**D**) We zoomed in one of the regions of the voltage-temperature curve where pronounced generation of the second harmonic (0.1 Hz) is observed, and showed it as a function of time to emphasize the modulation of the response. (**E**) The scheme of the experimental setup. A Hantek 2102B generator with internal impedance Z = 50 Ω was used to create a sinusoidal magnetic field in a solenoid "L". (**F**) The photo of the DAC H. (**G**) The photo of the DAC H in the solenoid before placing it in a liquid nitrogen dewar. The process of cooling and temperature stabilization was carried out in a nitrogen vapor.

Near the $T_C$, the polyhydride samples should contain many S-N-S contacts (N - normal metal, S - superconductor) in their composition due to the disordered structure and gradient of concentration of defects, hydrogen and, accordingly, the critical temperature. This was noted already in the very first years of research, for example in $YH_6$ [19,20]. In particular, it manifests in a small linear magnetoresistance [23,28], high upper critical magnetic fields $B_{C2}$ exceeding 200 T [24], broad SC transitions in ternary hydrides, very short electron mean free path [28], and other phenomena. In some situations, at a certain temperature close to $T_C$ and at a bias current close to the critical one $I_C$, the S-N-S contact arrays will randomly form SQUID-like loops and circuits connected in a standard 4-electrode scheme. In this case, there will be several SQUID loops of different sizes, $S_1$, $S_2$, $S_3$…, which will work simultaneously (inset of Fig. 2A, Figs. 4A-C). The maximum possible area of such loops is the area of the entire sample, which in our case (d ≈ 45 μm) is equal to $15.9\times10^{-10}$ m$^2$. One magnetic flux quantum $\Phi_0 = 2.068\times10^{-15}$ Wb will be achieved in the field of 1.3 μT = 0.013 G, which is approximately the sensitivity limit of the SQUID magnetometer in our experiments. By increasing the sample size to 100 μm, it is theoretically possible to achieve a sensitivity of ≈ 300 nT. Thus, to observe the SQUID effect in superhydrides, there is no need to create a special sample topology. This effect can be seen in many hydride samples prepared for transport measurements during cooling/warming in a weak modulated magnetic field (Fig. 1C, D). In the case of (La, Ce)$H_{10+x}$, we observed the appearance of complex oscillations of the resistance in certain temperature intervals near $T_C$ (Fig. 1D). Further analysis indicates the appearance of the SQUID effect in the sample (Fig. 2).

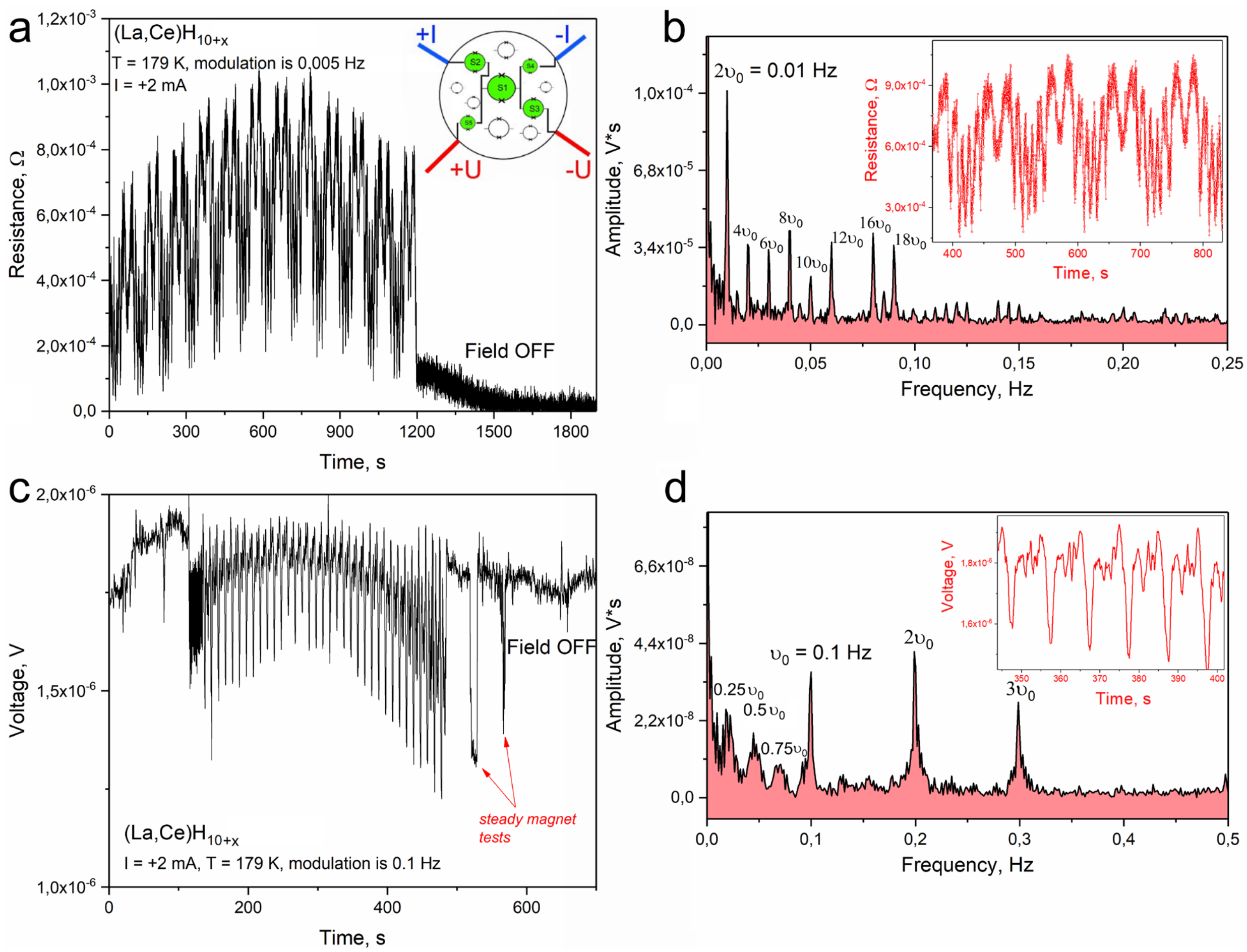

**Fig. 2. The DC SQUID effect and generation of higher harmonics in the (La, Ce)$H_{10+x}$ sample at the pressure of 148 GPa and the temperature of 179 ± 0.5 K.** The external magnetic field was generated by a sinusoidal current passing through a solenoid with a frequency of (**A**) 0.005 Hz and (**C**) 0.1 Hz. The bias current through the SQUID was +2 mA in both cases. Panel (**A**) shows the sample resistance measured in delta mode [29] with a period of 100 ms. The "Field OFF" zone corresponds to a turned off external magnetic field. The inset shows a qualitative drawing explaining the presence of multiple SQUID circuits ($S_1$, $S_2$, $S_3$…) in the sample in different

temperature/current intervals. The figures (**B**, **D**) show the Fourier analysis of the resulting complex periodic signal, which shows the presence of pronounced peaks of higher harmonics, characteristic of the SQUID effect in periodic magnetic fields. Insets: the analyzed signals *R(t)* and *U(t)*. The pronounced signal asymmetry in panels (**C**) and (**D**) is explained by the diode effect in the hydride sample [30]. Peaks at ~ 0.025,0.05, 0.075 Hz in Figure (**D**) correspond to non-integer flux quantization.

We examined synthesized superconducting (La,Ce)$H_{10+x}$ sample in the DAC H at 148 GPa and found that in a certain range of temperature 179 ± 0.5 K and bias current 1-2.5 mA, there is an unusually strong sensitivity of electrical resistance and the sample voltage drop to a weak low-frequency magnetic field of the solenoid "L" (Figures 1D, 2). A low frequency of 0.1-0.005 Hz was used to suppress inductive effects while keeping in mind also possible future applications in biology [31]. The observed SQUID effect does not depend on the direction of the DC current ($I_{\mathrm{bias}}$ = –2 mA, Supporting Figure S2) and is seen in delta mode resistance measurements as well (Fig. 2A) [29]. Direct measurements of the magnetic field of the solenoid show that in the mode of maximum signal amplitude (nominal peak-to-peak $\mathscr{E}_0$ = 20 V), the low-frequency signal generator allows creating a field of about $B_{max}$ = 5 Gauss in the solenoid, measured by a Hall sensor. The sensitivity of the DAC H and the estimated signal-to-noise ratio (Table S1), reaching $(S/N)_{\mathrm{max}} \approx$ 83, show that we can detect weaker magnetic fields of the order of 0.1 Gauss.

Distinctive feature of SQUID circuits is the extremely high sensitivity of their electrical resistance (or critical current) with respect to a weak magnetic field (*B*), and the intrinsic periodic dependence of measured voltage on the magnetic flux (within RSJ-model [32-34])

$$R = \frac{U}{I_{bias}} = \frac{R_J I_c}{I_{bias}} \sqrt{\left(\frac{I_{bias}}{2I_c}\right)^2 - \cos^2\left(\pi\frac{\Phi}{\Phi_0}\right)}, \qquad (1a)$$

where Φ - is a magnetic flux passing through a SQUID circuit [35], $R_J$ - is a resistance of the Josephson junction and $I_C$ – its critical current, while $I_{bias}$ – is a DC current used for measurements. Since the experiments were performed at the optimized current $I_{bias} = 2I_C = 2$ mA (Fig. 2A), where the modulation amplitude was greatest, the formula (1a) simplifies

$$R = \frac{U}{I_{bias}} = \frac{R_J}{2}\sin\left(\pi\frac{\Phi}{\Phi_0}\right). \qquad (1b)$$

Due to the modulation of the external magnetic field in our experiment $B(t) = B_{max}\sin(2\pi\upsilon t + \varphi_0)$. For simplicity, let us set the phase shift $\varphi_0$= 0, then, for small arguments of the sine function, we can use the Taylor series

$$R(t) = \frac{U(t)}{I_{bias}} \propto \sin\left(\frac{\pi\Phi_{\mathrm{max}}}{\Phi_0}\sin(2\pi\upsilon t)\right) = \left(\frac{\pi\Phi_{\mathrm{max}}}{\Phi_0}\sin(2\pi\upsilon t)\right) - \frac{1}{6}\left(\frac{\pi\Phi_{\mathrm{max}}}{\Phi_0}\sin(2\pi\upsilon t)\right)^3 + \frac{1}{120}\left(\frac{\pi\Phi_{\mathrm{max}}}{\Phi_0}\sin(2\pi\upsilon t)\right)^5 + \ldots, \qquad (2a)$$

where $\Phi_{\mathrm{max}}$ - a maximum magnetic flux passing through a SQUID circuit, ν - a modulating field frequency. If we assume that $\pi\Phi_{\mathrm{max}}/\Phi_0 = 1$, then the expression (2a) can be simplified further and we can understand where the higher harmonics in the frequency spectrum come from (Fig. 2B).

$$R(t) \propto 0.88\sin(2\pi\upsilon t) + 0.039\sin(6\pi\upsilon t) + 5.21 \cdot 10^{-4}\sin(10\pi\upsilon t) \ldots \qquad (2b)$$

Since we have asymmetric SQUID which also demonstrates pronounced diode effect (Supporting Figure S12), as well as non-zero phase shifts ($\varphi \neq 0$) and $\pi\Phi_{\mathrm{max}}/\Phi_0 \neq 1$ (Taylor expansion still converges), it becomes obvious that the experimental signal *U(t)* or *R(t)* can contain many different even and odd harmonics, which is indeed observed (see Fig. 2, Supporting Figures S2-S4, S13-S15). The occurrence of subharmonics (Fig. 2D, $\nu = {}^{1}/_{4}\ \nu_0, {}^{1}/_{2}\ \nu_0, {}^{3}/_{4}\ \nu_0$), corresponding to the non-integer flux quantization, is also noted in a number of cases and is often observed for inhomogeneous samples, in a series of connected SQUIDs, or if the SQUID lies on a metal plate [36,37].

The Little-Parks effect [38] may be another reason for observing small periodic oscillations in the resistance of a cylindrical superconductor. The amplitude of oscillations in the “dirty” limit is given by the formula [39]

$$\frac{\Delta T_C(B)}{T_C} = \frac{\Delta R}{(dR/dT)T_C} = 0.73\frac{\xi_0 l_e}{r^2}\left(n - \frac{\Phi_0}{\Phi}\right)^2, \qquad (3)$$

where $\Delta R$ – is amplitude of resistance oscillations, $dR/dT$ – is the slope of $R(T)$ function, $\xi_0$ – is the coherence length at 0 K, and $l_e$ – is the electron mean free path. For superhydrides, all of the listed characteristic lengths are on the order of nanometers: $\xi_0 \approx 1.5$ nm for $LaH_{10}$ [40], and $l_e \approx 1.6$ nm for similar ternary hydride $(La,Nd)H_{10}$ [28]. Considering that in our case $dR/dT \approx 0.013$ Ω/K, we come to the conclusion that the amplitude of the Little-Parks resistance oscillations is about several μΩ for micron loops ($r$), which is approximately 100-500 times less than the values observed in our experiment (Fig. 2A, 0.5-0.7 mΩ). Thus, the Little-Parks effect is too small to explain the observed strong resistivity oscillations in the $(La,Ce)H_{10+x}$ sample.

***AC measurements***

The presence of higher harmonics in the SQUID spectrum allows us to use the SR830 lock-in amplifier to measure the voltage drop across the sample in AC mode (Fig. 3A), for example, at the second harmonic, which is almost always the most pronounced. Indeed, at a current on the sample $I_{bias} = +2$ mA, pronounced generation of the 2nd harmonic is observed from the frequency of the modulating magnetic field ($\nu_0 = 50$ mHz, Fig. 3B). Peaks of 2nd harmonic generation occur at the temperature of N–S→ S–S contact transitions at the grain boundaries of various superconducting phases. At least two or three different superconducting phases with different $T_C$'s are present in the sample. These phases inevitably form during high-pressure and high-temperature synthesis of superhydrides in DACs and are possibly characterized by a different hydrogen content: for example, $(La, Ce)H_9$ and $(La, Ce)H_{10}$ [22,24,25].

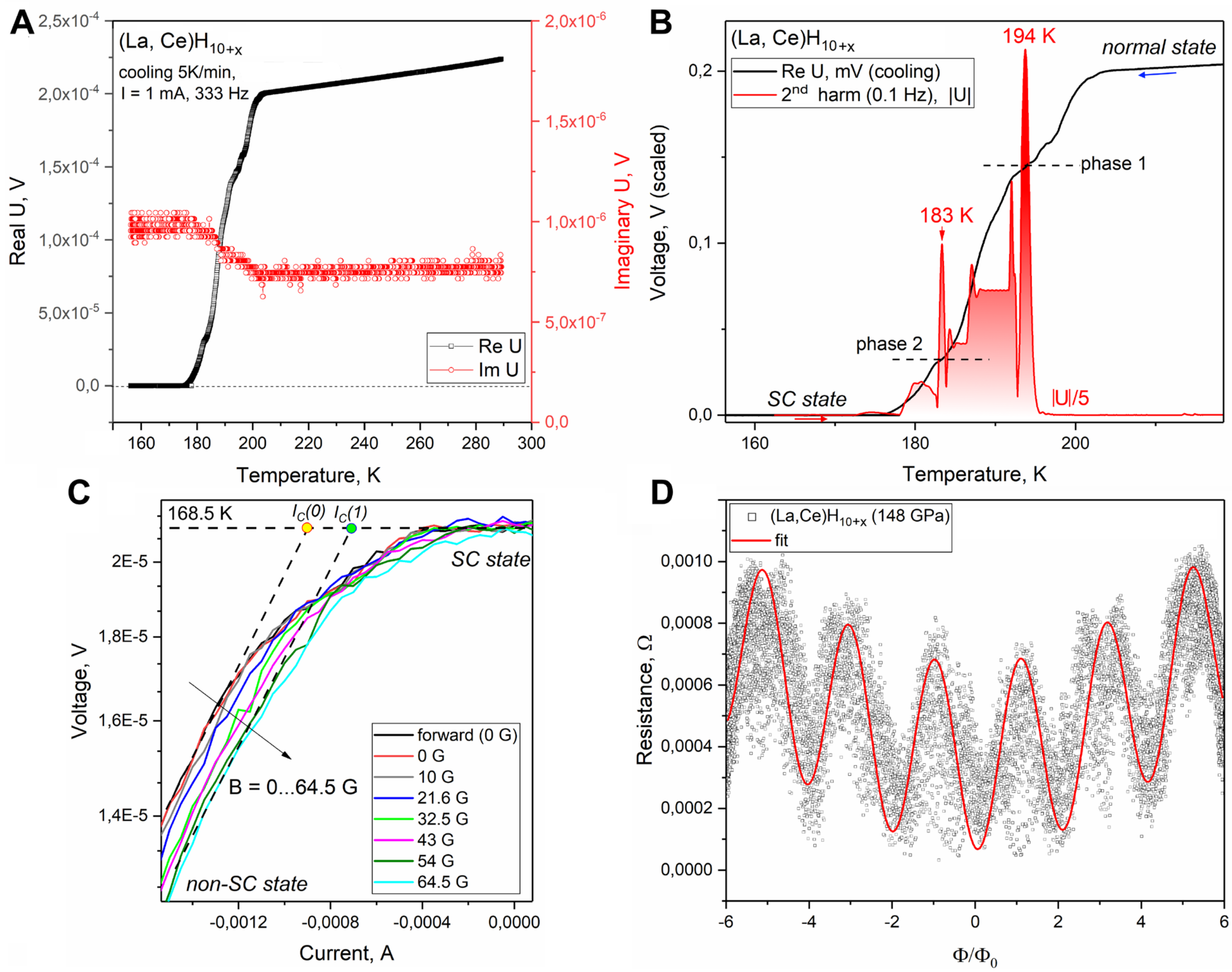

**Fig. 3. The AC transport study of the (La, Ce)$H_{10+x}$ sample in DAC H at 148 GPa in a modulated low-frequency magnetic field.** A four-electrode van der Pauw [41] circuit was used. (**A**) The temperature dependences of the real (Re U, black curve) and imaginary (Im U, red curve) components of the voltage on the sample. The cooling cycle with a rate 5 K/min in zero magnetic field. For measurements, a sinusoidal current of 1 mA (RMS) with a frequency of 333 Hz was used. (**B**) Absolute voltage ($|U|^2 = (ReU)^2 + (ImU)^2$) on the second harmonic ($2\nu_0$) of the external field modulation frequency $B_{max} \approx 5$ G, $\nu_0 = 0.05$ Hz (red curve) and its temperature dependence compared to the real AC voltage component taken from panel "a" (ReU, black curve). It can be seen that the signal peaks of the 2nd harmonic correspond to the "shelves" of the *U(T)* curve and probably relate to the changing nature of the grain boundaries: N-N → S-N → S-S. (**C**) Voltage-current characteristics of the sample at lower temperature, 168.5 K. $I_c(0)$ corresponds to the critical current in the absence of a magnetic field, whereas $I_c(1)$ is the critical current in the maximum field studied (64.5 G). The resulting *U-I* curves are similar to ones of a weak-link based YBCO SQUID magnetometer described in [42] (**D**) The dependence of the (La, Ce)$H_{10+x}$ SQUID resistance on magnetic flux Φ in the flux quantum $\Phi_0$ units calculated under the assumption of characteristic radius of the SQUID $r \sim 3$ μm. Raw data depending on the generator voltage are given in Supporting Figure S11.

In the vicinity of $T_C$(offset), at 180 K, the generation of the 2nd harmonic is associated with the SQUID effect in the sample (Fig. 3B), while the remaining signal is probably related to the diode effect accompanying asymmetrical SQUID circuits [30]. This explains the fact that in the normal and superconducting state, the generation of the 2nd harmonic is not observed despite the continued operation of the solenoid with the same frequency of 50 mHz. Therefore, sensing of a signal at the second harmonic can be used to detect SC transition, diode and SQUID effects in hydride superconductors.

The study of critical currents and *U-I* characteristics was carried out at temperatures significantly lower than 179 K in order to decrease electrical resistance of the sample, avoid problems with heating and to ensure that a significant part of the *U-I* characteristic was superconducting (flat). At temperatures below 179 K, the SQUID effect in the (La, Ce)$H_{10+x}$ sample is weakly expressed (Supporting Figure S8), which leads to a weak dependence of the critical current on the external field of the solenoid up to 64.5 G. Nevertheless, dependence of the critical current on the field is also observed at 168.5 K (Fig. 3C) providing another argument in favor of presence of Josephson junctions in the sample.

Finally, Fig. 3D and Supporting Figure S11 demonstrate the classical periodic dependence of the SQUID signal on the applied magnetic field in coordinates of the generator voltage connected to the solenoid or magnetic flux Φ. It is very similar to what was observed for the first YBCO ceramic SQUID in 1987 [43]. This relationship is based on the data in Fig. 2A. The signal amplitude changes due to heating and sample temperature drift when passing such a significant current (2 mA) through DC SQUID normal resistance of Josephson contacts $R_N/2 = 0.2 – 1$ mΩ at 179 K. The signal period corresponds to a magnetic field of about 1.7 – 1.8 G, and we can estimate the radius of the superconducting circuit as $r \sim 3$ μm.

**Discussion**

Inhomogeneities of the superconducting phase play an important role in the observed SQUID and diode effects. In particular, in the range of 194-200 K, no peculiarities in the behavior of the 2nd harmonic are observed, although part of the sample is in a superconducting state (Figs. 3B, C). Figs. 2, S1-S2 also show that the observed effect is caused by modulation of the external magnetic field of the solenoid and disappears when it is turned off (“Field OFF”). The DC SQUID effect exists in a wide range of modulation frequencies from 0.1 to 0.005 Hz, in a bias current range from 1.5 to 2.5 mA (Table S1), and can be reproduced in different superhydrides, for instance, in (La,Sc)$H_{12}$ [44] and in $La_4H_{23}$ [45]. In all cases, generation of higher harmonics up to $18\nu_0$ is found. The (La, Ce)$H_{10+x}$ sample exhibits high sensitivity to an external magnetic field of a permanent magnet located within 0.3-0.5 m from the sample (Fig. 2C, “steady magnet tests”).

Proposed idea of random formation of hydride rings during laser heating of ammonia borane in DACs can be illustrated by a relatively large and well-studied (50×50 map, 2 μm step) sample of Ba-Si polyhydrides (Figs. 4A-C). Since the synthesis in that symmetric electrode DAC with a wide opening angle was carried out by a similar way of pulsed laser heating of a BaSi sample and ammonia borane, we believe that the results in the case of La-Ce hydrides will be qualitatively identical. In Figs. 4B, C, characteristic round patterns of hydride phases with a diameter of up to 15 μm are clearly visible. They correspond to isothermal lines around the focal spot of IR laser used for the sample heating. Presence of such superconducting hydride rings of (La,Ce)$H_{10+x}$ causes the observed DC SQUID effect in the DAC H.

As previously mentioned, the asymmetric DAC H with d = 15.3 mm is intended for transport measurements only and cannot be investigated by X-ray methods. Therefore, we prepared another sample for XRD (Figs. 4D-F), synthesized at 152 GPa in a DAC with wide opening angle, using a similar alloy $La_{86}Ce_{14}$ and ammonia borane. As a result, we obtained a previously unknown La-Ce hydride, the diffraction pattern of which (Fig. 4D) can be explained by the hexagonal structure

*P*6$_3$/*mmc*-(La,Ce)$H_{12}$ with the unit cell parameters a = 4.816 Å, c = 3.815 Å, V = 38.32 Å$^3$/(La, Ce), which corresponds to a ratio of heavy atoms to hydrogen of 1:12. The unusual for *P*6$_3$/*mmc* ratio *c*/*a* = 0.79 indicates that the heavy atom sublattice may be described by a different hexagonal space group. It should be noted that hydrides with similar composition have been synthesized previously: these are $LaH_{12}$ [17] and (La,Sc)$H_{12}$ [44]. However, given the previous work on the La-Ce system, which led to the discovery of (La,Ce)$H_{9-10}$ [24], and considering high $T_C$(onset) > 200 K, in this paper we designate the studied compound as (La,Ce)$H_{10+x}$, where x can take values between –1 and +2.

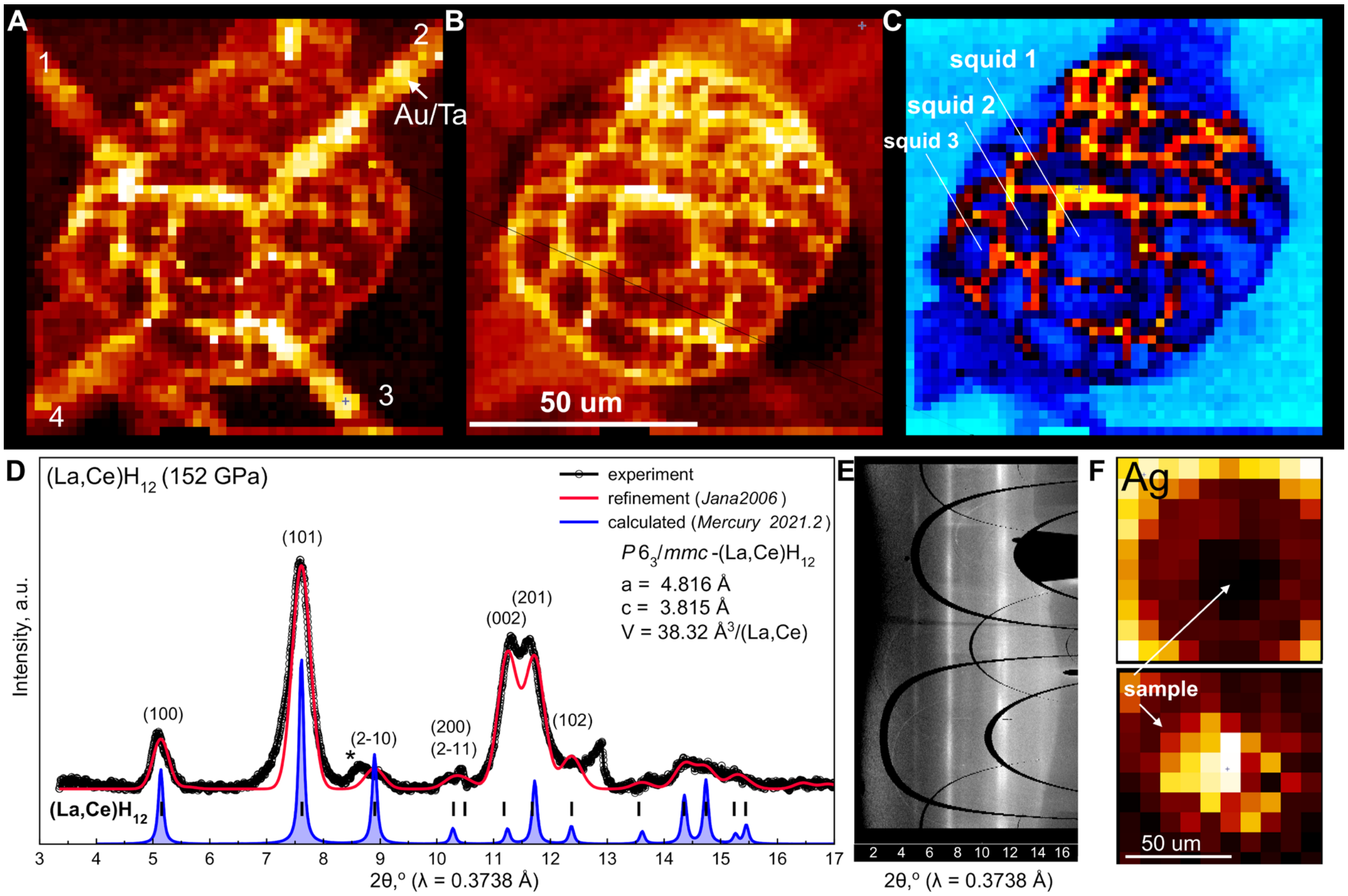

**Fig. 4. X-ray diffraction studies of polyhydrides in special DACs with a wide opening angle, synthesized via laser heating of metals and alloys with ammonia borane.** (A-C) Ba-Si polyhydrides at around 40 GPa. Au/Ta electrodes 1-4 can be clearly seen (A), as well as round patterns of hydride phases with weak links (B, C: "squid 1-3") formed along the isothermal lines near the laser focal spot. (D-E) X-ray diffraction (XRD) pattern of La-Ce hydride synthesized from $La_{86}Ce_{14}$ alloy and $NH_3BH_3$ at 152 GPa in a special XRD DAC. Powder X-ray diffraction analysis and the Le Bail refinements of the unit cell of (La,Ce)$H_{12}$ phase at 152 GPa. Black points mark experimental data, the red line – is the refinement, and the blue one – is predicted XRD of *P*6$_3$/*mmc*-(La,Ce)$H_{12}$. Asterisks mark unexplained reflections. (E) X-ray diffraction image ("cake"). (F) Material distribution map for composite gasket (Ag) and (La,Ce)$H_{12}$ hydride (sample).

One of the important parameters that describe the efficiency of a SQUID magnetometer is the so-called screening parameter $\beta_L = 2I_cL/\Phi_0$, where $L$ – the inductance of the system, $I_C$ – the Josephson junction critical current. The $\beta_L$ parameter for an optimized SQUID is usually of the order of unity [35]. Assuming the ideal case of a single-turn coil with a diameter of 45 μm and a height of 1 μm, we obtain the inductance $L_{max}$ = 70 – 100 pH and $\beta_L$ > 100. Moreover, if the SQUID circuit is smaller than the (La, Ce)$H_{10+x}$ sample diameter, then the inductance decreases significantly. So, if we have a ring with a radius of 3 μm, then $L \approx 10$ pH and $\beta_L \sim 10$. Not surprisingly, our "natural" DC SQUID turns out to be not the most optimal one due to the operating current being too high. The parameters found are close to those for YBCO SQUID with weak links (scratches on thin film) [46].

It is generally believed that for a SQUID to operate effectively, two basic conditions should be met [35]:

(1) Quantum interference dominates thermal fluctuations:

$$L < L_F = \frac{\Phi_0^2}{4\pi^2 k_B T} = 44\, pH\ at\ 179\ K, \qquad (4)$$

where $T$ – the operating temperature, $k_B$ – the Boltzmann constant. This condition is satisfied in our case for fairly small SQUID circuits of less than 20 μm, but it limits the possibilities for further increasing the sensitivity of superhydride SQUIDs due to their high operating temperature, which can already reach 240 K [47].

(2) The Josephson coupling energy should be greater than the thermal energy:

$$\Gamma = \frac{2\pi k_B T}{\Phi_0 I_0} = \frac{I_{th}}{I_0} = \frac{7.5\ \mu A}{I_0} \ll 1, \qquad (5)$$

where $I_{th}$ – the current of thermal fluctuations. Considering that the operating current $I_{bias}$ = 1 – 2.5 mA in our case, we come to the conclusion that the (La, Ce)$H_{10+x}$ DC SQUID operates in the mode of small thermal fluctuations and retains the potential to reduce the working current by 10-100 times. Despite the noise and heterogeneity of the described hydride SQUIDs, the possibility of their use in combination with compact thermoelectric coolers at temperatures of –25–30 °C, as we recently showed for (La,Sc)$H_{12}$ [44], opens ways to the creation of portable magnetometers on Li batteries for on-site field research.

**Conclusions**

We discovered the DC SQUID effect in lanthanum-cerium polyhydride (La, Ce)$H_{10+x}$ at temperature of 179 ± 0.5 K and pressure of 148 GPa, caused by a random inhomogeneous spatial distribution of superconducting gap in a microscopic sample. Periodic dependence of the sample resistance on the magnetic field allows to estimate the characteristic size of the SQUID circuit as $d$ ~ 6 μm, with a normal resistance of the Josephson "weak" links being about 0.2 – 1 mΩ at the operating temperature. Pronounced sample response to an external modulated magnetic field of ~ 5 G is observed when passing a direct bias current of 1 – 2.5 mA. The screening parameter $\beta_L = 2I_c L/\Phi_0 \sim 10$ in our device and can be further improved. Considering the signal-to-noise ratio $S/N$ up to 83, the DC SQUID based on a high-pressure diamond anvil cell DAC H can be used to detect weak magnetic fields down to 0.1 G. The sensitivity of superhydride SQUIDs of similar size can be improved at least by a factor about 10 to 0.01 G, and the operating temperature can reach 230-240 K [44].

Moreover, even more advanced applications could be envisioned by taking advantage of microfabrication tools to prepare more controllable and sophisticated superhydride-based SQUID devices. Such devices will certainly bring the quantum properties of SQUIDS close to room temperature, allowing the study of Josephson junction-based qubits in thermoelectrically cooled devices.

**Acknowledgments:** D.V.S. and D.Z. thank Beijing Natural Science Foundation (grant No. IS23017) and National Natural Science Foundation of China (NSFC, grant No. 12350410354) for support of this research. D. Z. thanks China Postdoctoral Science Foundation (Certificate No. 2023M740204) and finical support from HPSTAR. D.S. and D.Z. are grateful for financial support from the EMFL-ISABEL project. The high-pressure experiments were supported by the Russian Science Foundation (Project No. 22-12-00163). V.V.S acknowledges the financial support from Shanghai Key Laboratory of Material Frontiers Research in Extreme Environments (MFree), China (No. 22dz2260800) and Shanghai Science and Technology Committee, China (No. 22JC1410300). We thank Dr. Dominique Laniel and Dr. Umbertoluca Ranieri (University of Edinburgh, UK) for their help with XRD investigation of Ba-Si-H system.

**Author contributions:** D.S. and D.Z. conceived the original idea. D.S., D.Z., W. C. and I. T. performed experiments. D.S. and V.V.S. analyzed the data and wrote the manuscript. V.V.S. and H.K.M supervised the project. All the authors discussed the results and offered useful inputs.

**Competing interests:** Authors declare that they have no competing interests.

**Data and materials availability:** The authors declare that the main data supporting our findings of this study are contained within the paper and Supporting Information. All relevant data are available from the corresponding authors upon request. Raw data is available on the GitHub website: https://github.com/mark6871/LaCeH10/tree/SQUID.

### Supplementary Materials

Methods

Additional transport measurements in a periodic magnetic field

Current-voltage characteristics of (La, Ce)$H_{10+x}$

Simulation of higher harmonic generation in DC SQUID

# Supplementary Materials for

## Observation of the Josephson effect in superhydrides: DC SQUID based on $(La,Ce)H_{10+x}$ with operating temperature of 179 K

Dmitrii V. Semenok[1,†,*], Ivan A. Troyan[2,†], Di Zhou[1,†,*], Wuhao Chen[3,4] , Ho-kwang Mao [1] and Viktor V. Struzhkin[1,5,*]

[1] *Center for High Pressure Science & Technology Advanced Research, Bldg. #8E, ZPark, 10 Xibeiwang East Rd, Haidian District, Beijing, 100193, China*

[2]*A.V. Shubnikov Institute of Crystallography» of the Kurchatov Complex of Crystallography and Photonics (KKKiF), 59 LeninskyProspekt, Moscow 119333, Russia*

[3]*Quantum Science Center of Guangdong–Hong Kong–Macao Greater Bay Area (Guangdong), Shenzhen, China*

[4]*Department of Physics, Southern University of Science and Technology, Shenzhen 518055, China*

[5]*Shanghai Key Laboratory of Material Frontiers Research in Extreme Environments (MFree), Shanghai Advanced Research in Physical Sciences (SHARPS), Pudong, Shanghai 201203, China*

[†]These authors contributed equally to this work

Corresponding authors: Dmitrii Semenok (dmitrii.semenok@hpstar.ac.cn), Di Zhou (di.zhou@hpstar.ac.cn), and Viktor Struzhkin (viktor.struzhkin@hpstar.ac.cn)

## Content

## 1. Methods

To load the high-pressure diamond anvil cell DAC H, consisting of a non-magnetic body made of 40HNU-VI alloy, diamond anvils with a culet diameter of 50 μm and sputtered Ta/Au electrical contacts, we used a particle of $La_4Ce$ alloy with a thickness of 1 μm and a diameter of 45 μm. The $La_4Ce$ alloy was prepared according to previously described procedure [24]. Ammonia borane ($NH_3BH_3$, AB) was used as a hydrogen source. After laser heating, which consisted of several ~0.3 s pulses of IR laser (wavelength is 1.04 μm), the pressure in DAC H increased from 137 to 148 GPa.

The sample temperature was stabilized in the range of 179 ± 0.5 K using a Lakeshore 335 temperature controller and insulated NiCr wire with a thickness of 50 μm and a resistance of about 100 Ω wound around the diamond anvil cell DAC H (Figure 1f). The current passing through this heater created an additional magnetic field, which should be attributed to interfering noise. Cernox 1030 thermometer (error is ±0.2 K) was glued by $Al_2O_3$-based heat conductive adhesive on a DAC's rhenium gasket within 1 mm from the sample.

The magnetic field was generated using a Hantek signal generator (peak-to-peak amplitude was 10 or 20 V, maximum field in the solenoid ≈ 5 G). The generator signal was also used as a reference for the SR830 lock-in amplifier (Stanford Research). Lock-in amplifier parameters: time constant 30 ms, sensitivity 200 mV, detection period: 100 ms. We used a frequency of 50 mHz and a second harmonic was 100 mHz. For DC measurements we used Keithley 6221 current source and Keithley 2182A nanovoltmeter, detection period was 100 ms. Sample resistance measurements were carried out in the delta mode. Note that the delta mode measurements are performed at a minimum of two time points and then averaged. Because of this, the time axis is compressed by half (2t → t), which leads to a formal doubling of all frequencies (ν → 2ν) in the spectrum (for example, Figs. 2A, B).

**Table S1.** The experimental parameters used to study the $(La,Ce)H_{10+x}$ DC SQUID. The operating temperature was 179 ± 0.5 K in all cases.

| Bias current, mA | Modulation frequency, Hz | Signal/noise ratio (S/N) |
|---|---|---|
| +2 | 0.05 | - |
| +2 | 0.005 | 14 |
| +2.5 | 0.01 | 83 |
| +2 | 0.1 | 9.6 |
| +1.5 | 0.01 | 4.8 |
| 2* | 0.005 | 50 |
| –2 | 0.01 | 31 |

*Delta mode measurements of electrical resistance

## 2. Additional transport measurements in a periodic magnetic field

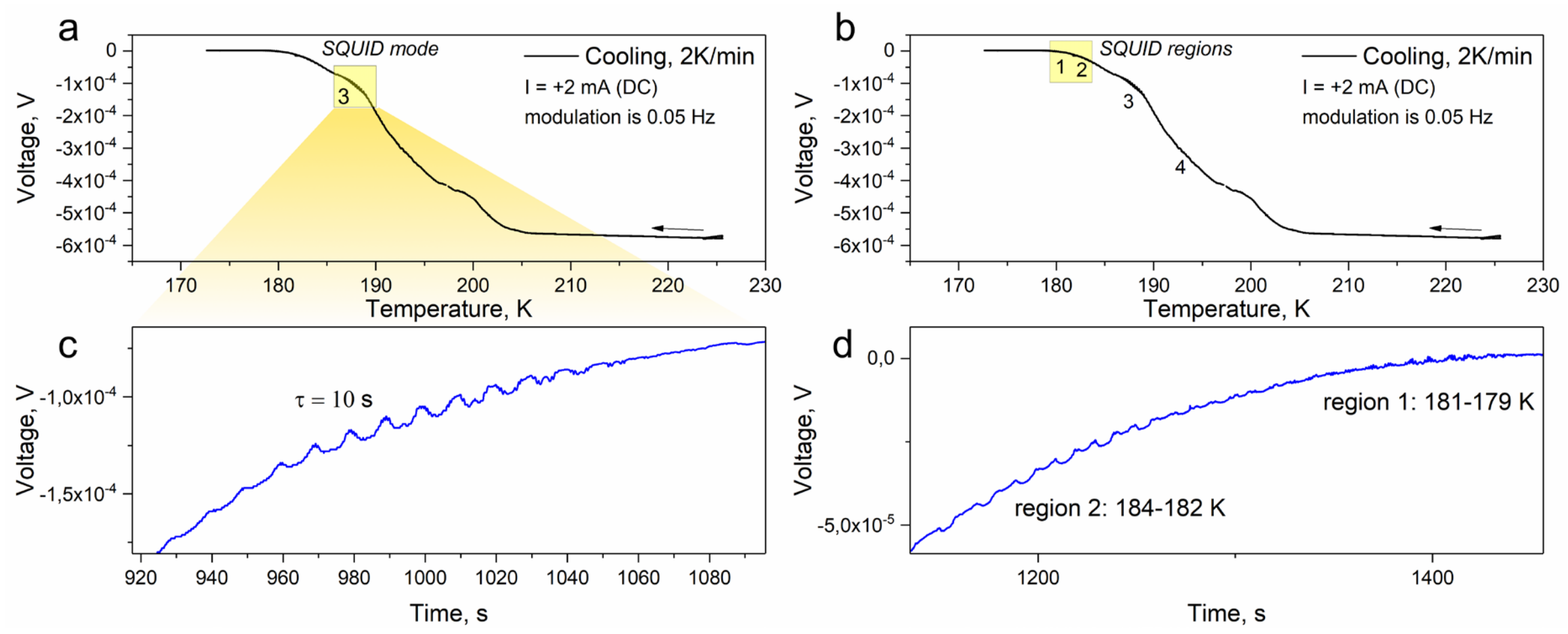

**Figure S1.** Dependence of the voltage drop across the sample in DAC H on temperature during the cooling cycle (rate is about 2 K/min) in a modulated magnetic field ($\nu_0$ = 0.05 Hz) with a direct current of 2 mA. The width of the superconducting transition is $\Delta T_C \approx 25$ K. Numbers 1-4 correspond to the temperature intervals of high sensitivity to the magnetic field and generation of higher harmonics. Intervals with high normal resistance ($R_N$) are inconvenient for transport studies due to significant heating of the sample by the bias current and drift of its temperature. (a, c) Interval "3" about 190 K and its zoom in. There is a pronounced generation of the 2nd harmonic ($2\nu_0$ = 0.1 Hz) relative to the frequency of the external magnetic field ($\nu_0$ = 0.05 Hz). (b, d) Intervals "1" and "2" and their zoom in. There is a noticeable generation of higher harmonics.

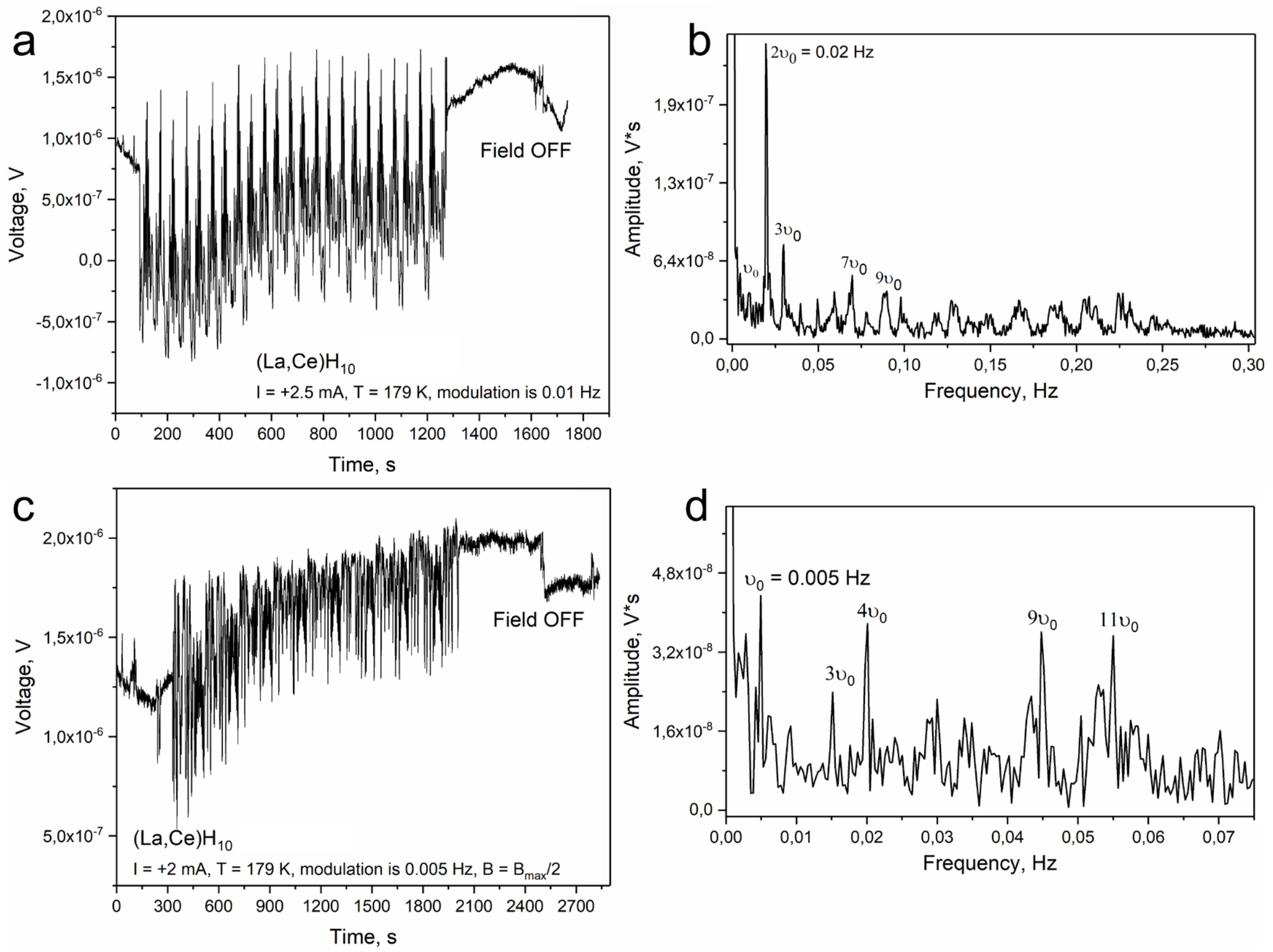

**Figure S2.** The DC SQUID effect and generation of higher harmonics in a (La, Ce)$H_{10}$ sample at pressure of 148 GPa and temperature of 179 K. An external magnetic field was generated by a sinusoidal current (Hantek 2102B generator) with a frequency of (a) 0.01 Hz and (c) 0.005 Hz. In the latter case, we used half the peak-to-peak amplitude of the signal (10 V instead of 20 V, $B_{max} \approx 2.5$ G) supplied to the solenoid. The bias DC current through the sample was equal to (a) + 2.5 mA and (c) +2 mA. The "Field OFF" region corresponds to a turned off external magnetic field. Figures (b, c) show the Fourier analysis of the resulting complex periodic signal, which shows the presence of pronounced peaks of higher harmonics, characteristic of the SQUID effect.

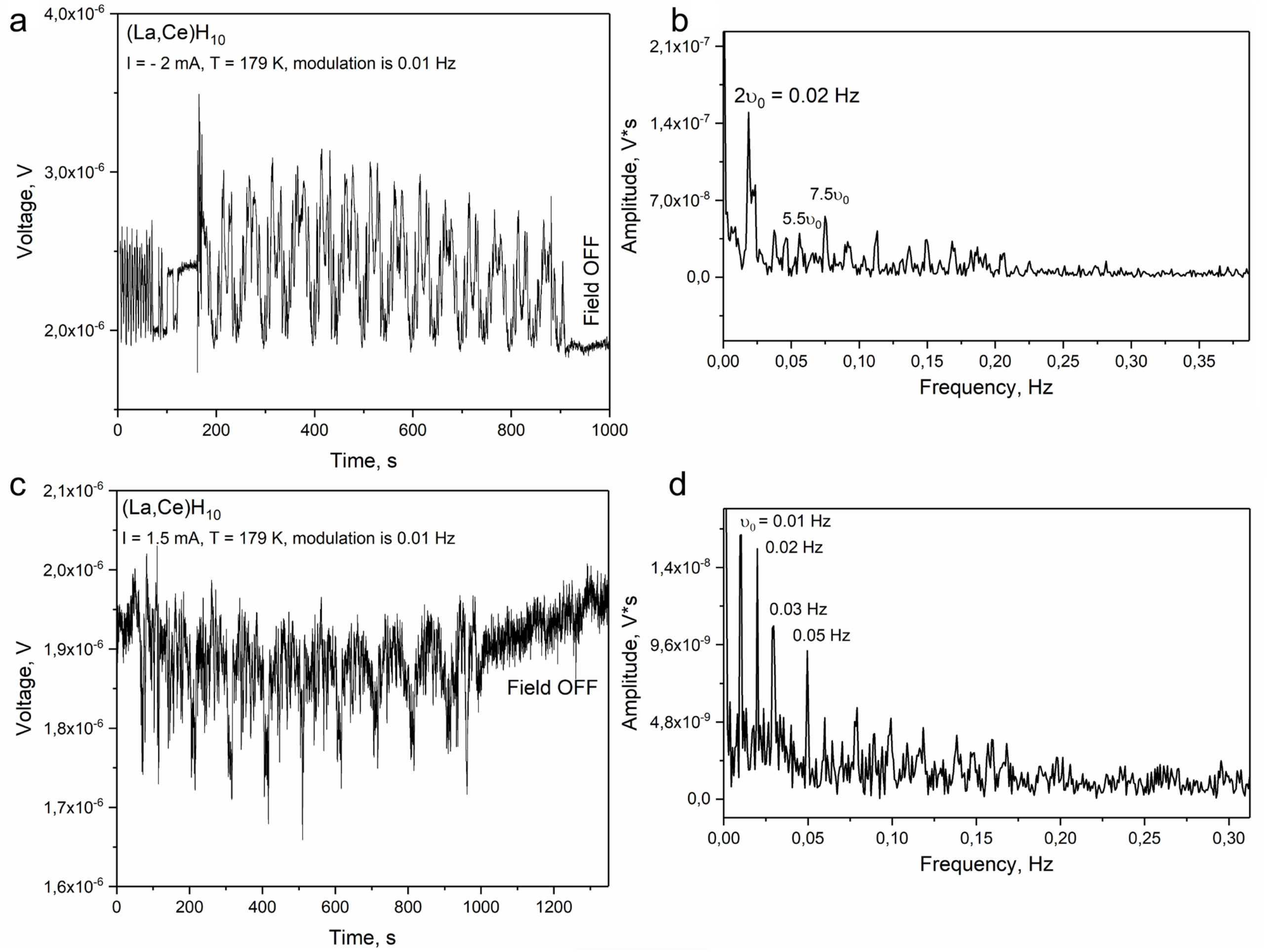

**Figure S3.** The DC SQUID effect and generation of higher harmonics in the (La, Ce)$H_{10+x}$ sample at pressure of 148 GPa and temperature of 179 K. The external magnetic field was generated by a sinusoidal current with a frequency of 0.01 Hz. The DC bias current through the sample was equal to (a) – 2 mA and (c) +1.5 mA. The "Field OFF" region corresponds to the turned off external magnetic field. Panels (b, c) show the Fourier analysis of the resulting complex periodic signal, which demonstrates the presence of pronounced peaks of higher harmonics, characteristic of the SQUID effect.

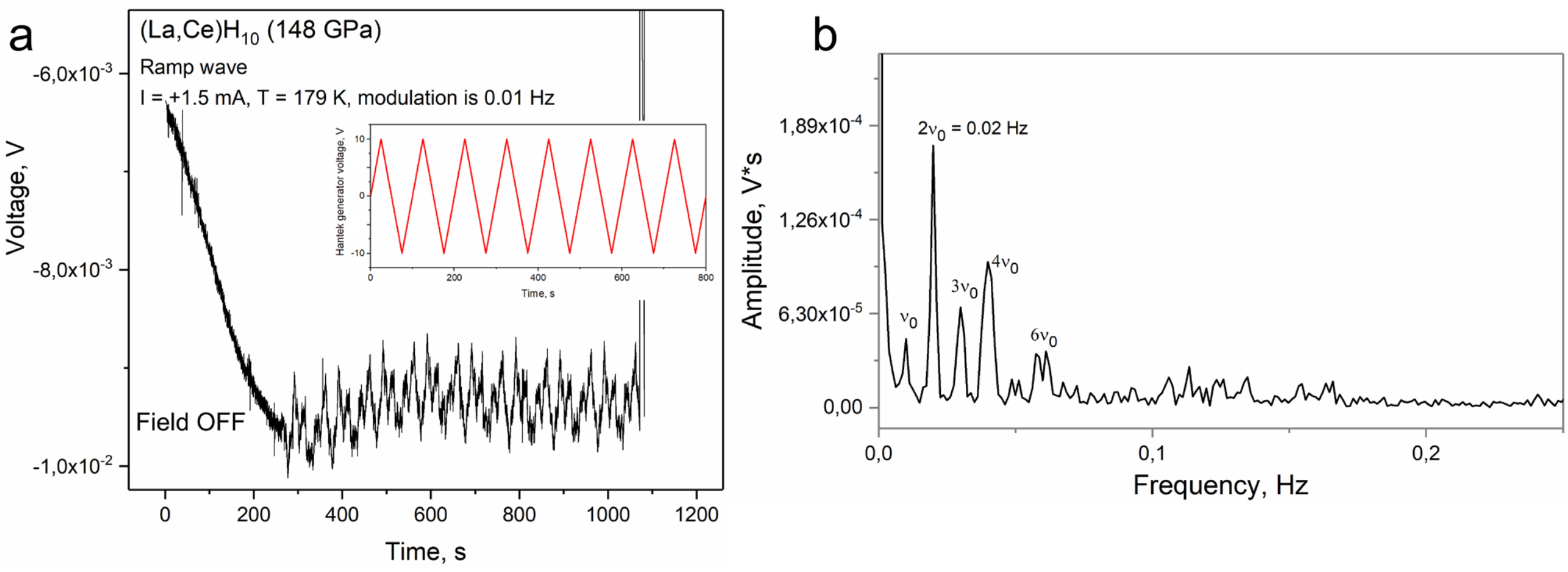

**Figure S4.** The DC SQUID effect and generation of higher harmonics in the (La, Ce)$H_{10+x}$ sample at pressure of 148 GPa and temperature of 179 K. The external magnetic field was generated by a triangular (ramp) wave current with a frequency of 0.01 Hz (panel "a": inset). The DC bias current through the sample was equal +1.5 mA. The "Field OFF" region corresponds to the turned off external magnetic field. Panel (b) shows the Fourier analysis of the resulting complex periodic signal, which demonstrates the presence of pronounced peaks of higher harmonics, characteristic of the DC SQUID effect.

### 3. Current-voltage characteristics of (La, Ce)$H_{10+x}$

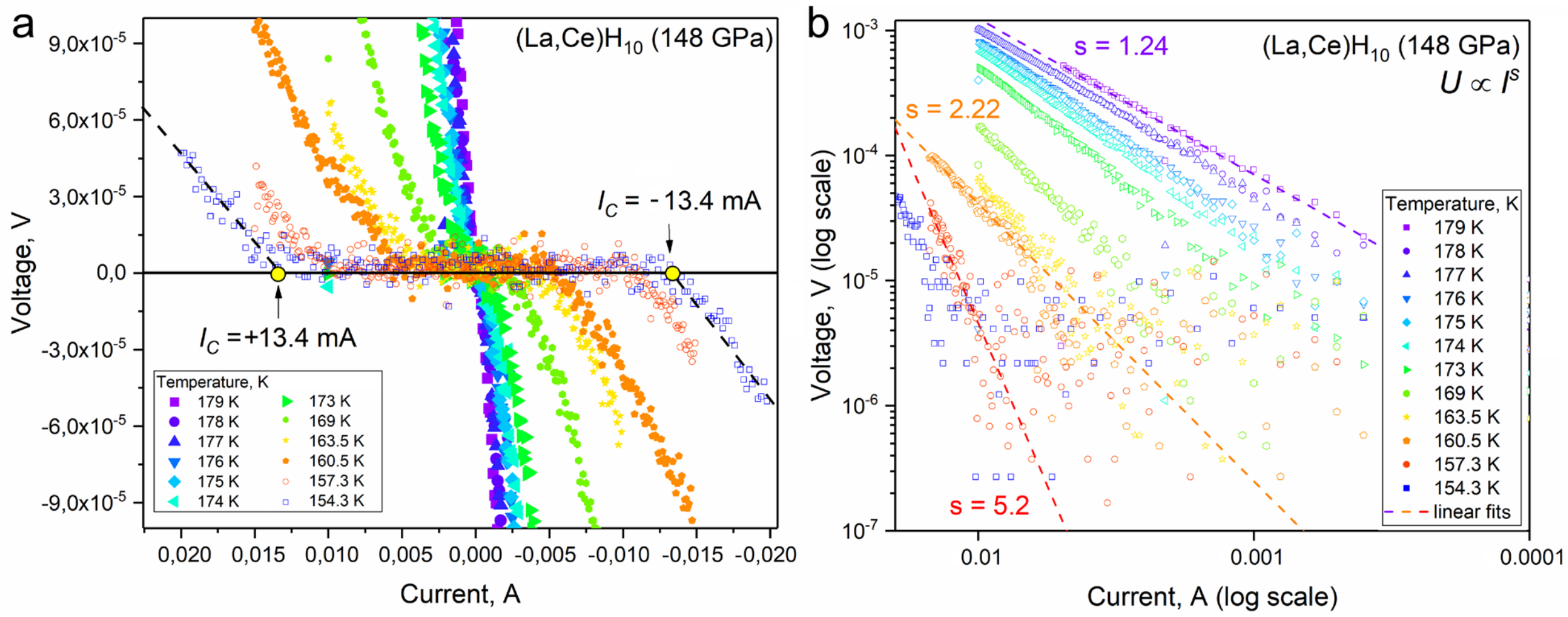

**Figure S5.** Current-voltage characteristics of the (La, Ce)$H_{10+x}$ sample measured at 148 GPa in the range of 154.3 – 179 K. A Keithley 6221 current source was used with a step of 0.1-0.2 mA and a measurement time of 25 ms. (a) Evolution of the symmetrical current-voltage characteristics of the sample with decreasing temperature. At 154.3 K, the critical current reaches 13.4 mA. The diode effect is practically not expressed in this temperature range. (b) Power-law dependence of the sample voltage on the current, characteristic of many polycrystalline superconducting materials [46].

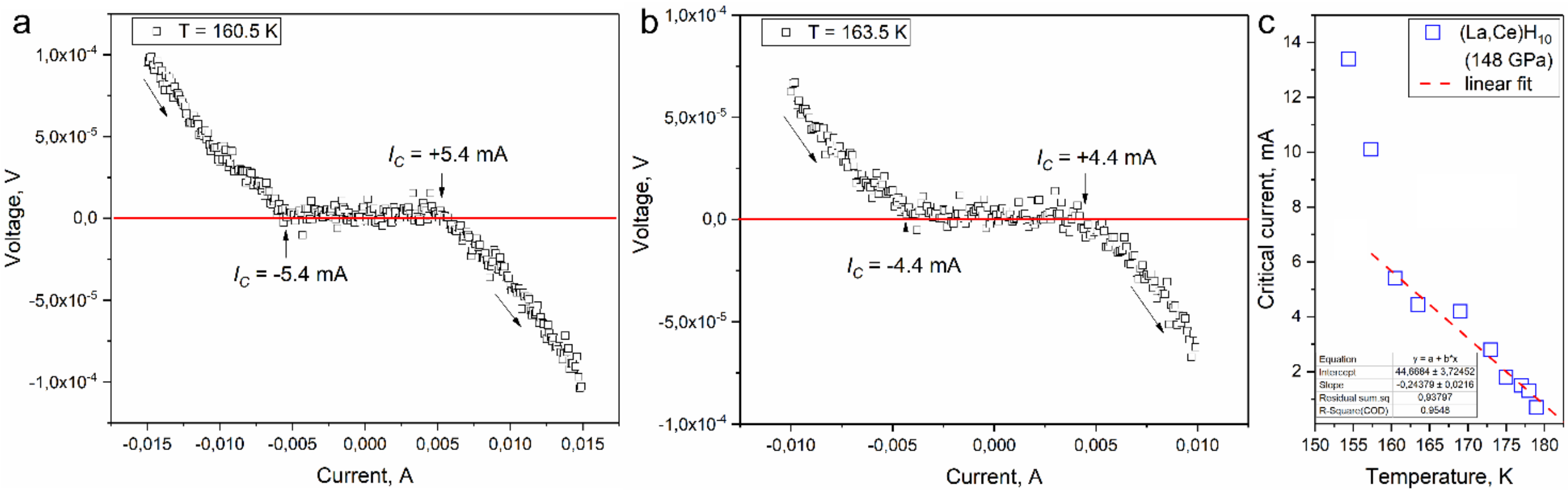

**Figure S6.** Examples of current-voltage characteristics of (La, Ce)$H_{10+x}$ taken at a temperature below the offset superconducting transition ($T$ < 179 K). (a) Current-voltage characteristics at 160.5 K. (b) Current-voltage characteristics at 163.5 K. The arrows indicate the direction of current change. (c) Dependence of critical current $I_C$ on temperature in (La, Ce)$H_{10+x}$. Near $T_C$ the dependence $I_C(T)$ is linear.

**Table S2.** Critical currents of (La, Ce)$H_{10}$ in a zero external magnetic field at different temperatures.

| Temperature, K | Critical current ($I_C$), mA | Temperature, K | Critical current ($I_C$), mA |
|---|---|---|---|
| 154.3 | 13.4 | 175 | 1.8 |
| 157.3 | 10.1 | 177 | 1.5 |
| 160.5 | 5.4 | 178 | 1.3 |
| 163.5 | 4.44 | 179 | 0.7 |
| 169 | 4.2 | | |
| 173 | 2.8 | | |

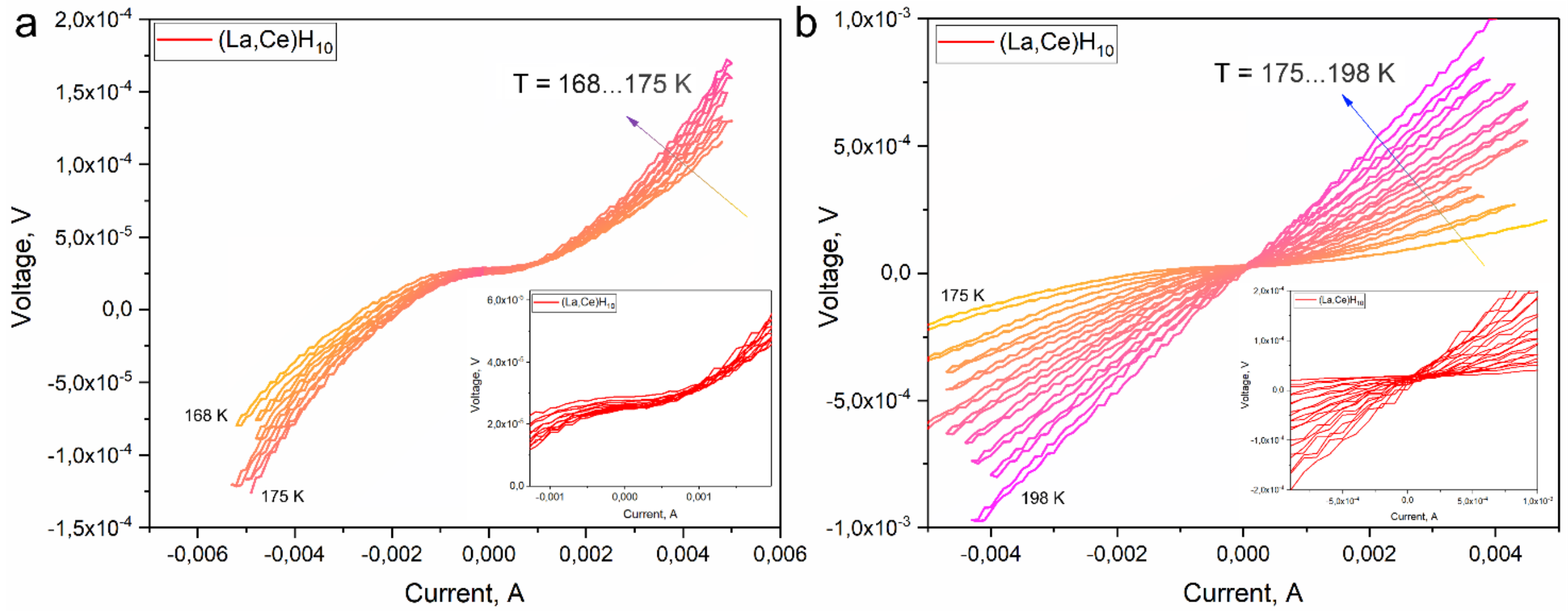

**Figure S7.** Current-voltage characteristics of (La, Ce)$H_{10+x}$ taken in the sweep mode over a continuously increasing temperature (a) from 168 to 175 K, (b) from 175 to 198 K. There is a certain voltage offset that can be caused by thermo electromotive force (EMF) in the sample. Due to the continuous change in temperature and the finite scan time, the "forward" and "back" current cycles are not the same.

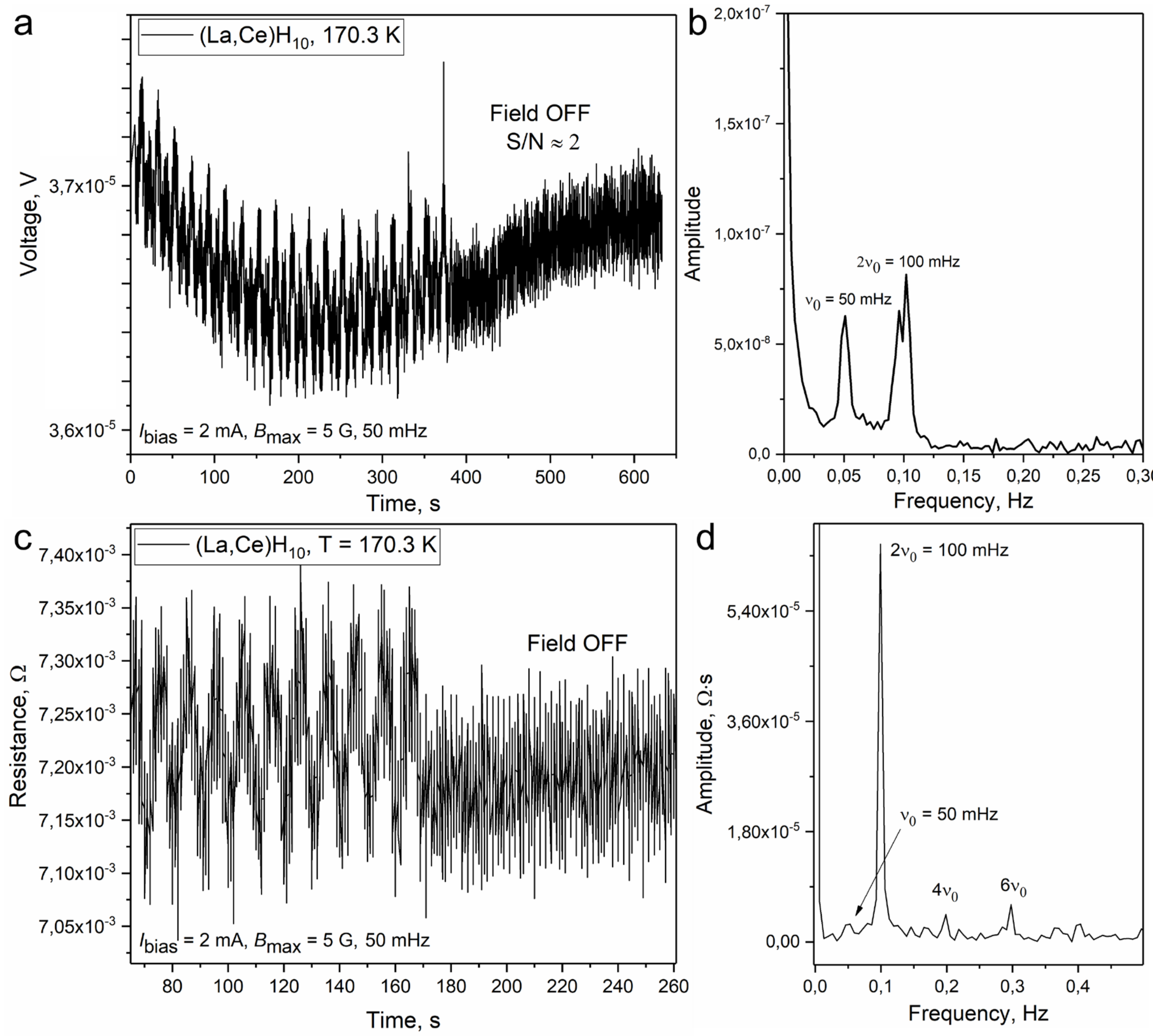

**Figure S8.** The DC SQUID effect and generation of higher harmonics in a (La, Ce)$H_{10+x}$ sample at pressure of 148 GPa and temperature of 170.3 ± 0.5 K. An external magnetic field was generated by a sinusoidal current of a Hantek 2102B with a frequency of 0.005 Hz. The peak-to-peak amplitude of the signal sent to the solenoid was 20 V ($B_{max} \approx 5$ G). The bias DC current through the sample was equal to 2 mA. The "Field OFF" region corresponds to a turned off external magnetic field. Figure (a) corresponds to the detected voltage drop across the sample, while Figure (c) shows the sample resistance, measured in the delta mode within an independent run. Figures (b, d) show the Fourier analysis of the resulting complex periodic signal, which shows the presence of pronounced peaks of higher harmonics.

There is one important special case of the DC SQUID effect in superhydrides, associated with the microscopic size of the superconducting loop that accidentally formed in the sample. If $\frac{\pi\Phi_{max}}{\Phi_0} \ll 1$, then the expansion in the formulas (1-2) can be stopped near the first term. In this case, we will see only several first harmonics in the frequency spectrum of such a "degenerate" DC SQUID. It is realized in experiments at 168.5 K and 170.3 K (see, for example, Figure S8b).

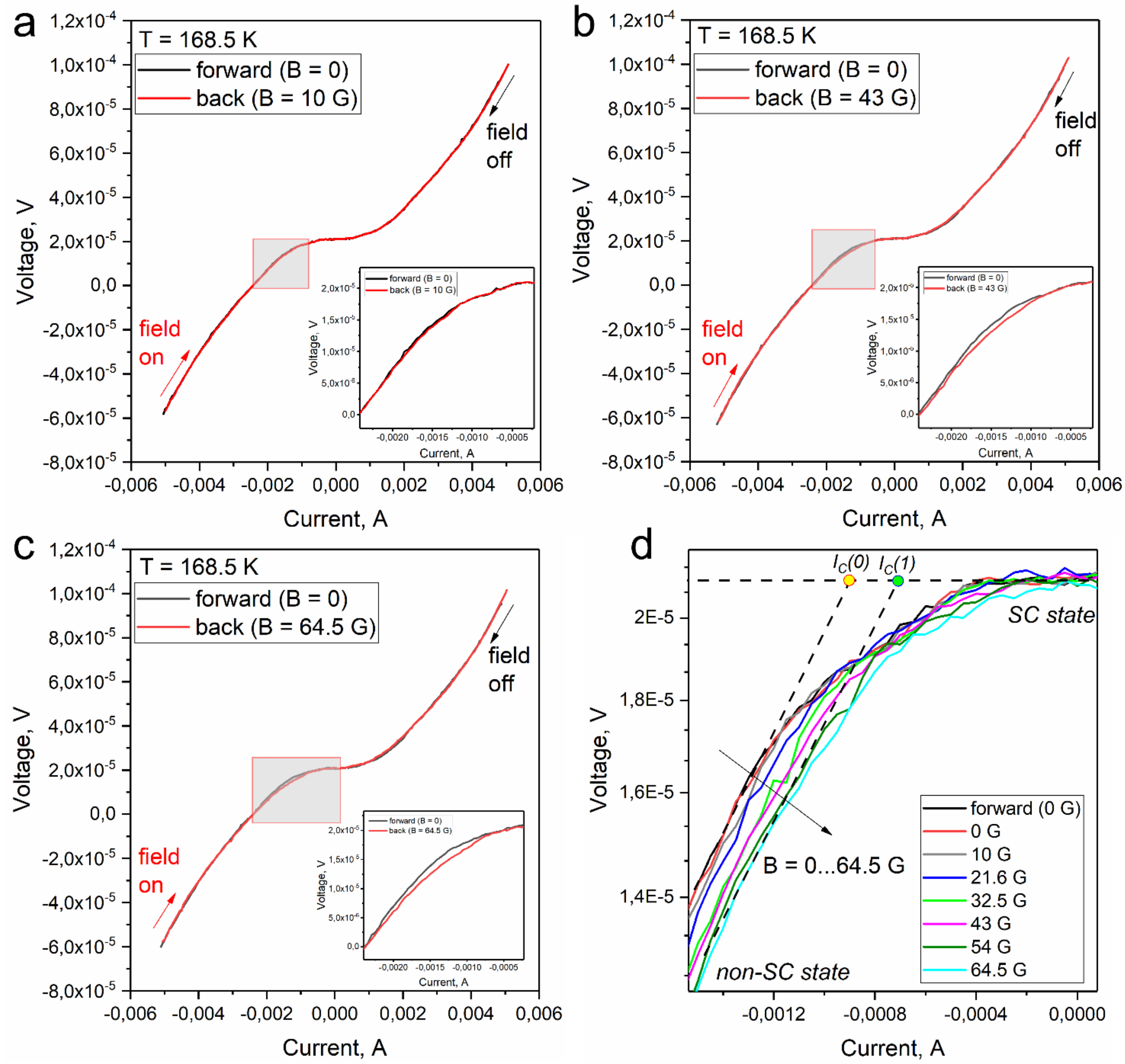

**Figure S9.** Current-voltage characteristics of (La, Ce)$H_{10+x}$ at 168.5 ± 0.5 K in a constant magnetic field. There is a certain voltage offset that can be caused by thermo EMF in the sample. *U-I* measurements were carried out in the absence ("forward scan" or "forward") and in the presence ("backward scan" or "back") of a constant magnetic field with an induction of up to 64.5 Gauss. Panel (a) shows that there is practically no difference between the curves when comparing *U-I* at zero field and at 10 G. Panel (b) indicates that with an increase in the field induction to 43 G, a noticeable decrease in the in-field critical current is observed. The in-field critical current decreases (panel (c)) with an increase of the field to 64.5 G. (d) Zoom in the region of transition from the superconducting (SC) state to the non-superconducting state in the vicinity of $I = 1$ mA. $I_c(0)$ corresponds to the critical current in the absence of a magnetic field, whereas $I_c(1)$ is the critical current in the maximum field studied. The resulting *U-I* curves are similar to ones of a weak-link based YBCO SQUID magnetometer described in Ref. [42].

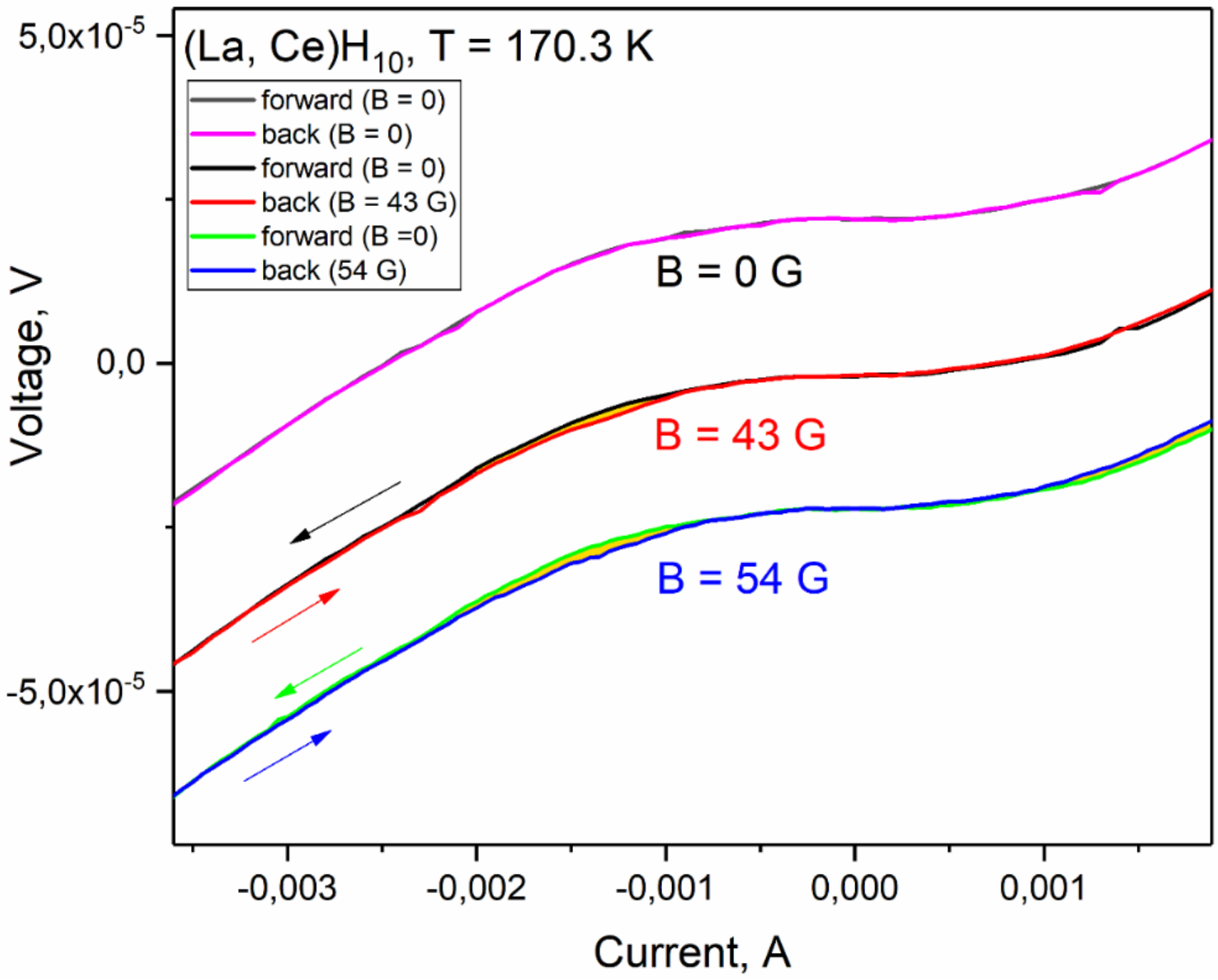

**Figure S10.** Current-voltage characteristics of (La,Ce)$H_{10+x}$ at 170.3 ± 0.5 K in a constant magnetic field. The curves were shifted vertically for ease of comparison. *U-I* measurements were carried out in the absence ("forward scan" or "forward") and in the presence ("backward scan" or "back") of a constant magnetic field with an induction of up to 54 Gauss. Upper curves show that there is practically no difference between the curves when comparing "forward" and "back" *U-I* at zero field. Middle curves indicate that with an increase in the field induction to 43 G, a noticeable difference between zero-field *U-I* and in-field *U-I* is observed (marked by yellow color). The same is for the lowest curves at 54 G.

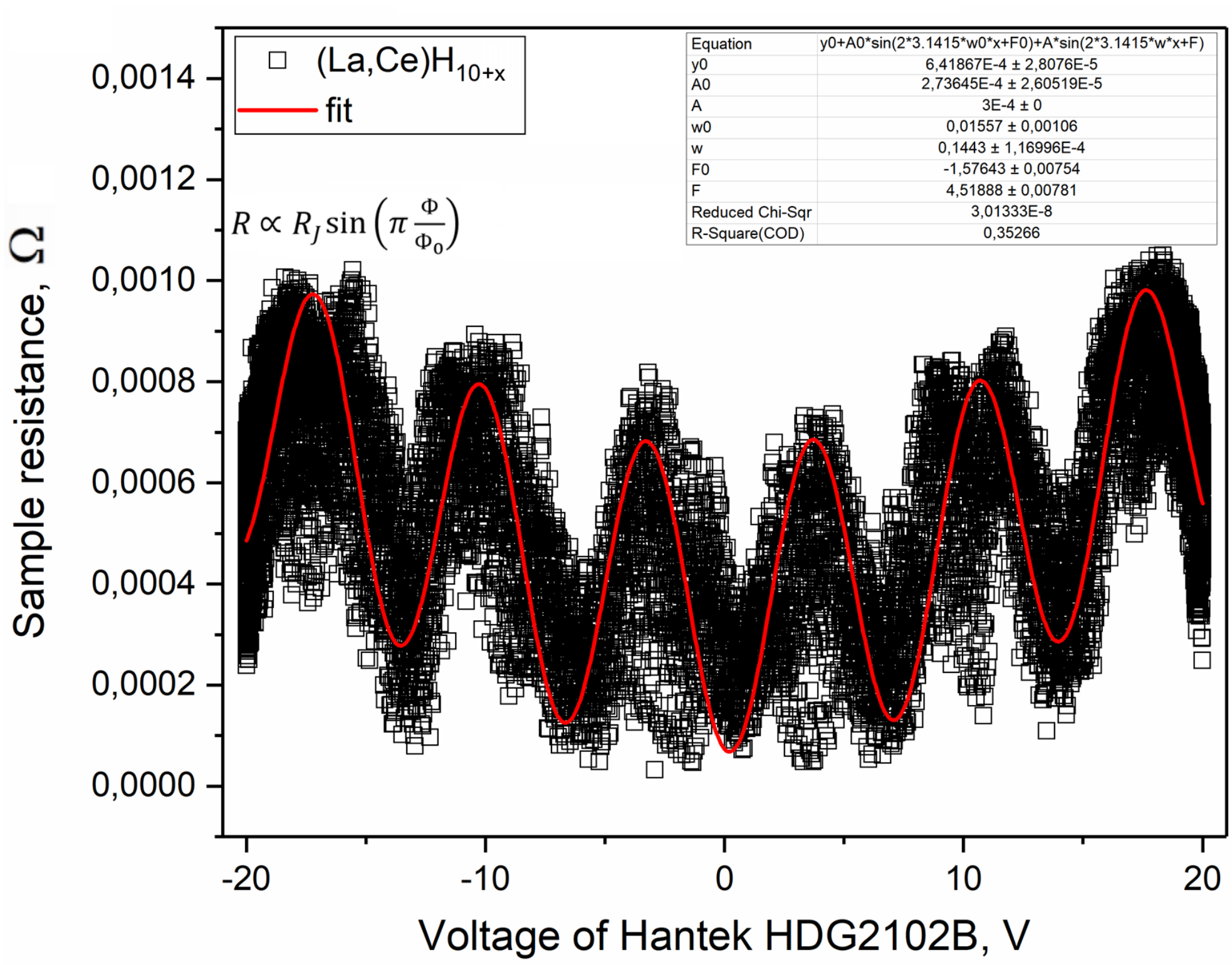

**Figure S11.** The dependence of the (La, Ce)$H_{10+x}$ DC SQUID resistance on the peak-to-peak voltage on solenoid connected with the Hantek HDG2102B generator. The peak-to-peak voltage of 20 V corresponds to the magnetic field of about ~ 5 G.

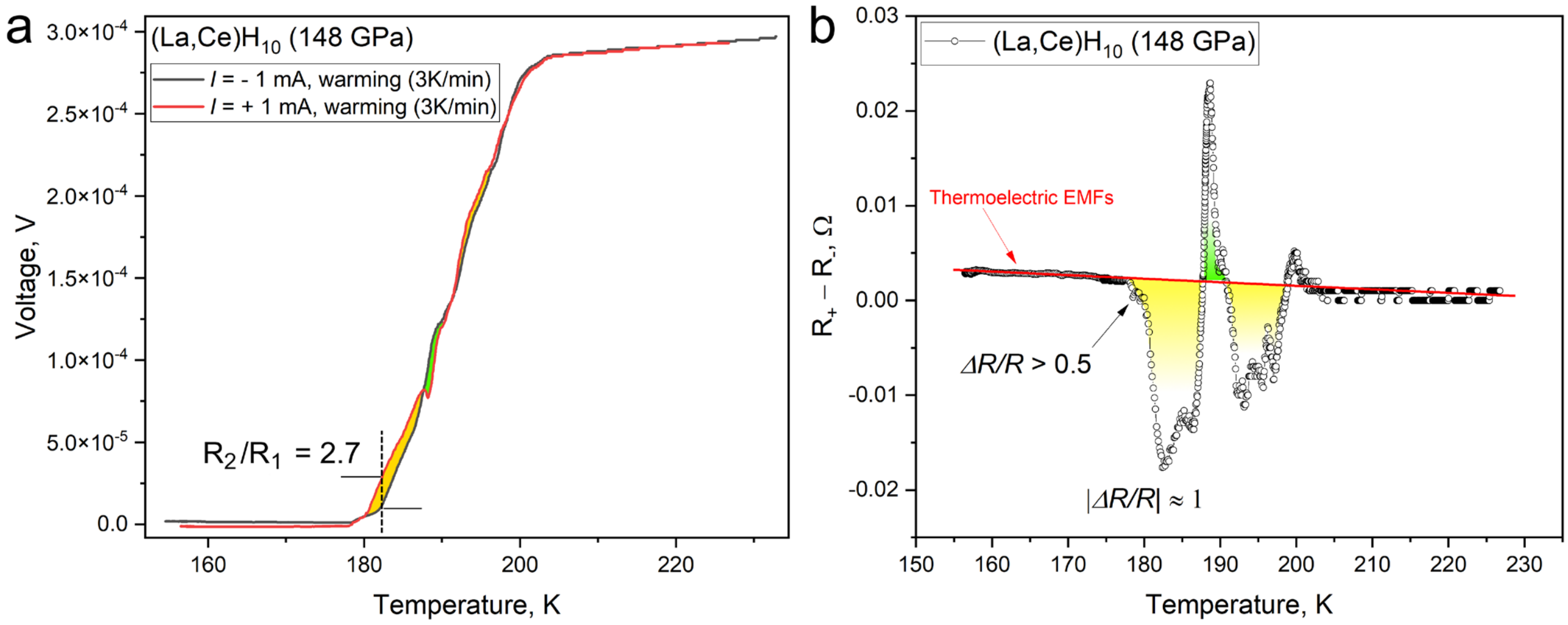

**Figure S12.** Diode effect in (La,Ce)$H_{10+x}$ at 148 GPa. (a) Dependence of the voltage drop across the sample in the DC measurement mode with different constant current directions (± 1 mA) in a warming cycle at a rate of 3 K/min. It can be seen that the resistance depends on the direction of the current and the ratio $R_+/R_-$ reaches 2.7 around 182-183 K. (b) Dependence of the difference of resistances $R_+ - R_-$ for current of different directions (± 1 mA) on temperature. The red straight line corresponds to the linear change in thermoelectric voltage (EMF) in the region under study. It is clearly seen that the sample demonstrates a pronounced asymmetry in conductivity.

## 4. Simulation of higher harmonic generation in DC SQUID

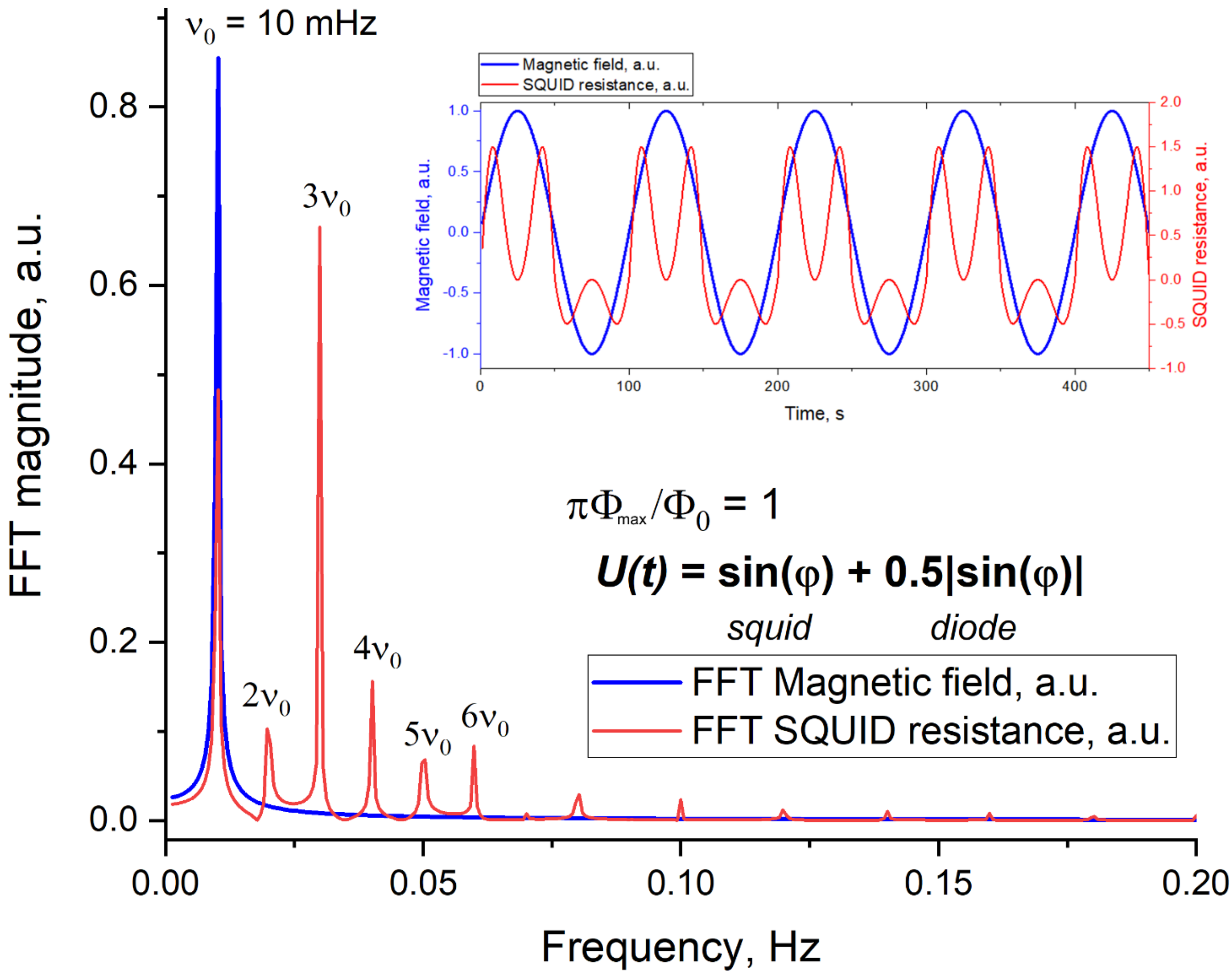

**Figure S13.** Modelling the frequency response of asymmetric SQUID in a sinusoidal magnetic field with a frequency of 10 mHz at the values $\Phi_{max}/\Phi_0 = 1$ with a 50 % "diode" contribution represented as the absolute value of the modulated current in the sample. Here $\varphi = \pi(\Phi_{max}/\Phi_0)\times\sin(2\pi\nu_0 t)$. The SQUID is weakly sensitive in this case. There are only a few harmonics in the frequency spectrum (red curve). Inset: the original signal (blue curve) from the solenoid and the electrical voltage on the sample as a function of time.

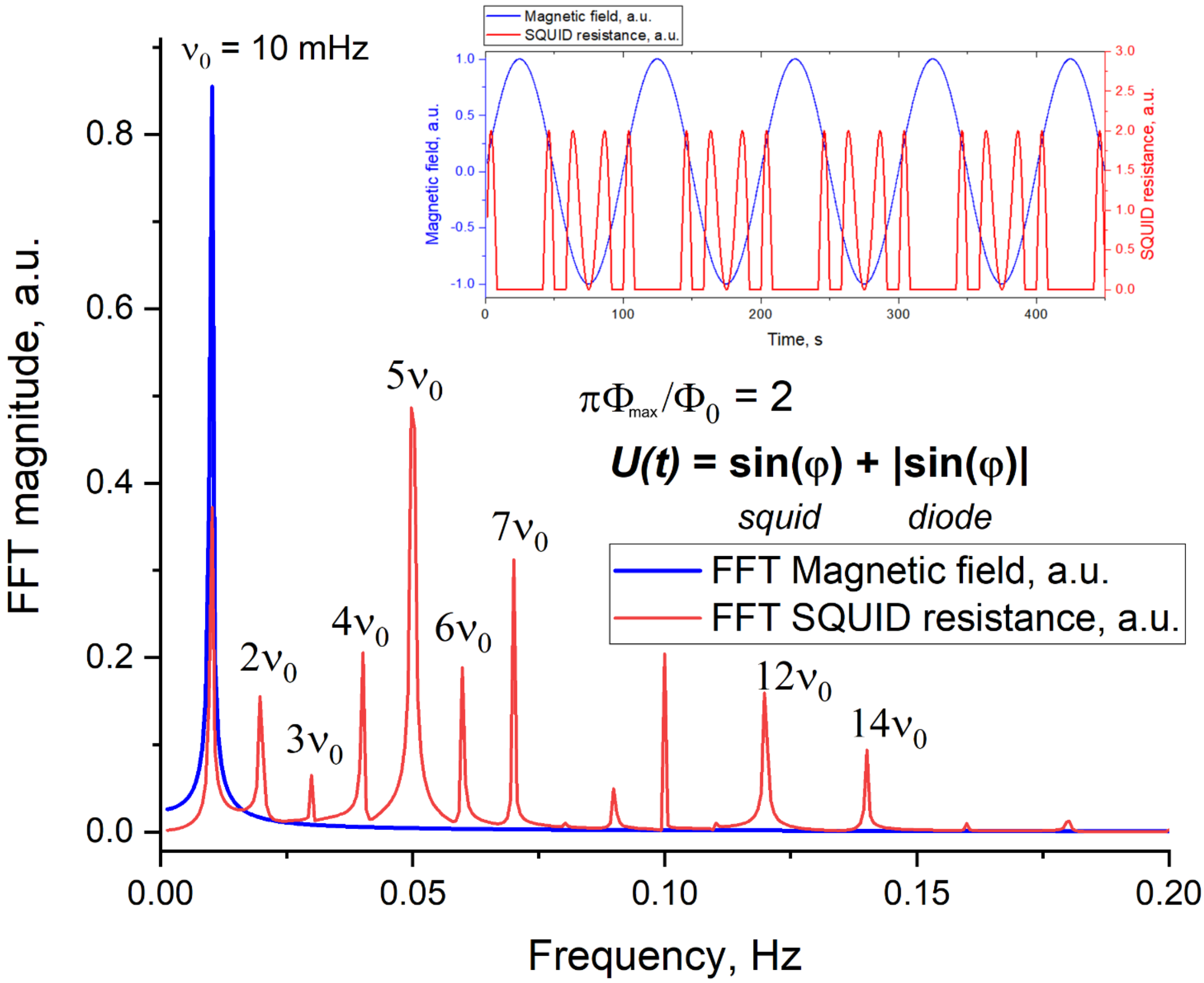

**Figure S14.** Modelling the frequency response of asymmetric SQUID in a sinusoidal magnetic field with a frequency of 10 mHz at the values $\Phi_{max}/\Phi_0 = 2$ with a 100 % “diode” contribution represented as the absolute value of the modulated current in the sample. Here $\varphi = \pi(\Phi_{max}/\Phi_0)\times\sin(2\pi\nu_0 t)$. There are many harmonics in the frequency spectrum (red curve). Inset: the original signal (blue curve) from the solenoid and the electrical resistance of the SQUID as a function of time.

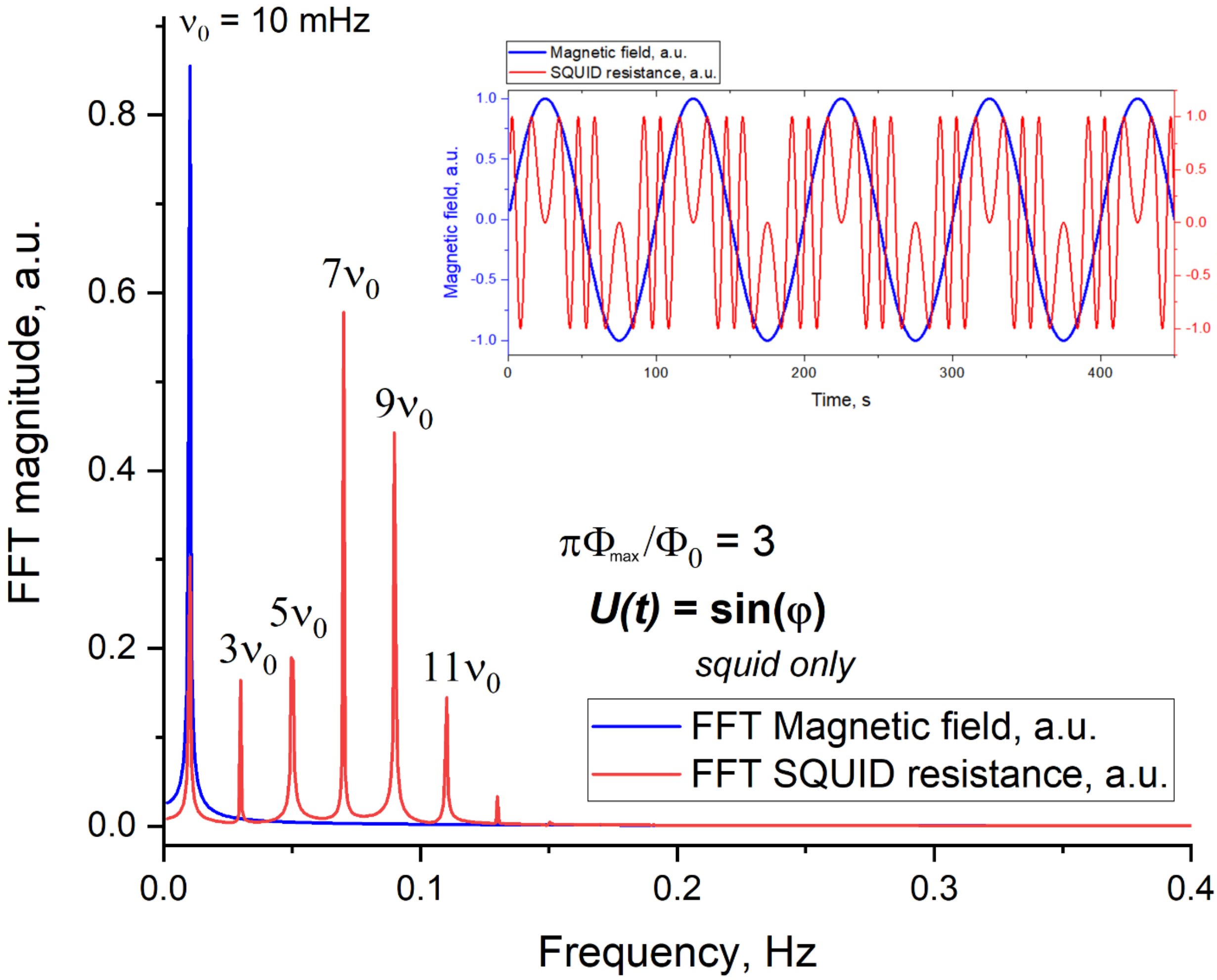

**Figure S15.** Modelling the frequency response of asymmetric SQUID in a sinusoidal magnetic field with a frequency of 10 mHz at the values $\Phi_{max}/\Phi_0 = 3$ with a 0 % "diode" contribution (symmetric SQUID). Here $\varphi = \pi(\Phi_{max}/\Phi_0)\times\sin(2\pi\nu_0 t)$. There are many odd harmonics in the frequency spectrum (red curve) which turn into even when resistance is measured in the delta mode. Inset: the original signal (blue curve) from the solenoid and the electrical resistance of the SQUID as a function of time.